\documentclass[longauth]{aa} 
\usepackage{xcolor}
\usepackage{support-caption}
\usepackage{subcaption}
\usepackage{graphicx}
\usepackage{threeparttable}
\usepackage{float}
\usepackage{booktabs}
\usepackage{longtable}
\usepackage{lscape}

\newcommand{\ha}{\ifmmode {\rm H}\alpha \else H$\alpha$\fi}
\newcommand{\hb}{\ifmmode {\rm H}\beta \else H$\beta$\fi}
\newcommand{\lya}{\ifmmode {\rm Ly}\alpha \else Ly$\alpha$\fi}
\newcommand{\pg}{\ifmmode {\rm P}\gamma \else Pa$\gamma$\fi}
\newcommand{\lyb}{\ifmmode {\rm Ly}\beta \else Ly$\beta$\fi}
\newcommand{\lyg}{\ifmmode {\rm Ly}\gamma \else Ly$\gamma$\fi}

\newcommand{\flyc}{\ifmmode \mathrm{f}_\mathrm{esc}\mathrm{(LyC)} \else $\mathrm{f}_\mathrm{esc}\mathrm{(LyC)}$\fi}

\def\kmsmpc{km s$^{-1}$ Mpc$^{-1}$}

\def\ergs{\ifmmode \mathrm{erg\hspace{1mm}s}^{-1} \else erg s$^{-1}$\fi}

\def\micron{\ifmmode \mu\mathrm{m} \else $\mu$m\fi}
\def\msun{\ifmmode \mathrm{M}_{\odot} \else M$_{\odot}$\fi}
\def\msunyr{\ifmmode \mathrm{M}_{\odot} \hspace{1mm}{\rm yr}^{-1} \else $\mathrm{M}_{\odot}$ yr$^{-1}$\fi}
\def\zsun{\ifmmode Z_{\odot} \else Z$_{\odot}$\fi}
\def\lsun{\ifmmode L_{\odot} \else L$_{\odot}$\fi}
\def\mstar{\ifmmode \mathrm{M}_{\star} \else M$_{\star}$\fi}
\newcommand{\hst}{HST}
\newcommand{\jwst}{JWST}

\newcommand{\NIRSpec}{NIRSpec}
\newcommand{\NIRCam}{NIRCam}

\usepackage{txfonts}
\newcommand{\orcid}[1]{\href{https://orcid.org/#1}{\includegraphics[width=10pt]{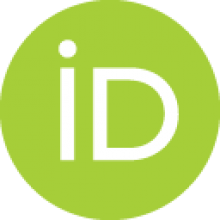}}}
\usepackage{hyperref}
\hypersetup{colorlinks, 
 linkcolor = blue,
 urlcolor = blue,
 citecolor = blue,
 anchorcolor = blue
}
\begin{document}

\title{The gradual decline of Ly$\alpha$ visibility in the CANDELS fields: evidence for the combined effects of galaxy evolution and reionization}

\titlerunning{The gradual decline of Ly$\alpha$ visibility at 4 < z < 14}
\authorrunning{L. Napolitano et al.}

 \subtitle{}
     \author{L. Napolitano \orcid{0000-0002-8951-4408}
 \inst{1}
 \and L. Pentericci \orcid{0000-0001-8940-6768}
 \inst{1}
 \and A. Ferrara \orcid{0000-0002-9400-7312}
 \inst{2}
 \and M. Llerena \orcid{0000-0003-1354-4296}
 \inst{1}
 \and M. Dickinson \orcid{0000-0001-5414-5131}
 \inst{3}
 \and A. Calabrò \orcid{0000-0003-2536-1614}
 \inst{1}
 \and \\ S. L. Finkelstein \orcid{0000-0001-8519-1130}
 \inst{4,21}
 \and R. Begley \orcid{0000-0003-0629-8074}
 \inst{5,6}
 \and P. Arrabal Haro \orcid{0000-0002-7959-8783}
 \inst{22, 7}
 \and A. Arroyo-Polonio \orcid{0000-0002-9523-8016}
 \inst{1}
 \and B. E. Backhaus \orcid{0000-0001-8534-7502}
 \inst{8}
 \and \\ D. Bevacqua \orcid{0000-0001-8863-2472}
 \inst{1}
 \and A. Bhagwat \orcid{0000-0003-0275-5506}
 \inst{9}
 \and M. Bischetti \orcid{0000-0002-4314-021X}
 \inst{10,11}
 \and M. Castellano \orcid{0000-0001-9875-8263}
 \inst{1}
 \and S.-J. Chang \orcid{0000-0002-0112-5900}
 \inst{9}
 \and E. R. Cueto \orcid{0009-0000-2942-6740}
 \inst{9}
 \and \\ V. D'Odorico \orcid{0000-0003-3693-3091}
 \inst{11,12}
 \and C. T. Donnan \orcid{0000-0002-7622-0208}
 \inst{3}
 \and M. Galbiati \orcid{0000-0002-1843-1699}
 \inst{11}
 \and G. Gandolfi \orcid{0000-0003-3248-5666}
 \inst{1}
 \and M. Giavalisco \orcid{0000-0002-7831-8751}
 \inst{13}
 \and M. Hirschmann \orcid{0000-0002-3301-3321}
 \inst{11,14}
 \and \\ J. Kartaltepe \orcid{0000-0001-9187-3605}
 \inst{15}
 \and A. M. Koekemoer \orcid{0000-0002-6610-2048} 
 \inst{16}
 \and K. K. Knudsen \orcid{0000-0002-7821-8873}
 \inst{17}
 \and R. A. Lucas \orcid{0000-0003-1581-7825}
 \inst{16}
 \and S. Mascia \orcid{0000-0002-9572-7813}
 \inst{18}
 \and L. Paquereau \orcid{0000-0003-2397-0360}
 \inst{17}
 \and \\ B. P\'{e}rez-D\'{\i}az \orcid{0000-0002-0939-9156}
 \inst{1}
 \and C. Piscitelli \orcid{0009-0004-3816-0656}
 \inst{11, 19}
 \and R. Rana \orcid{0000-0002-4213-407X}
 \inst{17}
 \and P. Santini \orcid{0000-0002-9334-8705}
 \inst{1}
 \and A. J. Taylor \orcid{0000-0003-1282-7454}
 \inst{4,21}
 \and E. Taylor \orcid{0000-0001-8728-2984}
 \inst{6}
 \and R. Tripodi \orcid{0000-0002-9909-3491}
 \inst{1,12}
 \and \\ S. M. Wilkins \orcid{0000-0003-3903-6935}
 \inst{20}
 \and L. Y. A. Yung \orcid{0000-0003-3466-035X}
 \inst{16}
 }
 \institute{\textit{INAF – Osservatorio Astronomico di Roma, via Frascati 33, 00078, Monteporzio Catone, Italy} 
 \email{lorenzo.napolitano@inaf.it}
 \and
 \textit{Scuola Normale Superiore, Piazza dei Cavalieri 7, 50126 Pisa, Italy}
 \and
 \textit{NSF's National Optical-Infrared Astronomy Research Laboratory, 950 N. Cherry Ave., Tucson, AZ 85719, USA} 
 \and
 \textit{Department of Astronomy, The University of Texas at Austin, Austin, TX, USA} 
 \and
 \textit{Armagh Observatory and Planetarium, College Hill, Armagh, BT61 9DG, N. Ireland, UK} 
 \and 
 \textit{Institute for Astronomy, University of Edinburgh, Royal Observatory, Edinburgh, EH9 3HJ, UK} 
 \and
 \textit{Astrophysics Science Division, NASA Goddard Space Flight Center, 8800 Greenbelt Rd, Greenbelt, MD 20771, USA} 
 \and
 \textit{Department of Physics, 196A Auditorium Road, Unit 3046, University of Connecticut, Storrs, CT 06269, USA} 
 \and 
 \textit{Max Planck Institut für Astrophysik, Karl Schwarzschild Straße 1, D-85741 Garching, Germany} 
 \and
 \textit{Dipartimento di Fisica "Enrico Fermi,", Università di Pisa, Largo Bruno Pontecorvo 3, Pisa I-56127, Italy} 
 \and 
 \textit{INAF - Osservatorio Astronomico di Trieste, Via G. B. Tiepolo 11, I-34131 Trieste, Italy} 
 \and
 \textit{IFPU - Institute for Fundamental Physics of the Universe, via Beirut 2, I-34151 Trieste, Italy} 
 \and
 \textit{University of Massachusetts Amherst, 710 North Pleasant Street, Amherst, MA 01003-9305, USA} 
 \and
 \textit{Institute of Physics, Laboratory of Galaxy Evolution, Ecole Polytechnique Federale de Lausanne (EPFL), Observatoire de Sauverny, 1290 Versoix, Switzerland} 
 \and
 \textit{Laboratory for Multiwavelength Astrophysics, School of Physics and Astronomy, Rochester Institute of Technology, 84 Lomb Memorial Drive, Rochester, NY 14623, USA} 
 \and
 \textit{Space Telescope Science Institute, 3700 San Martin Drive, Baltimore, MD 21218, USA} 
 \and
 \textit{Department of Physics and Astronomy, Chalmers University of Technology, SE-412 96, Gothenburg, Sweden} 
 \and
 \textit{Institute of Science and Technology Austria (ISTA), Am Campus 1, A-3400 Klosterneuburg, Austria} 
 \and
 \textit{Department of Physics, Astronomy Section, University of Trieste, Via G.B. Tiepolo, 11, I-34143 Trieste, Italy} 
 \and
 \textit{Astronomy Centre, University of Sussex, Falmer, Brighton BN1 9QH, UK} 
 \and 
 \textit{Cosmic Frontier Center, The University of Texas at Austin, Austin, TX, USA} 
 \and
 \textit{Center for Space Sciences and Technology, UMBC, 5523 Research Park Dr, Baltimore, MD 21228 USA} 
 }

\date{Received ... / Accepted ...}

\abstract{We investigate the evolution of \lya\ visibility and the physical properties of \lya\ emitters (LAEs) across the five CANDELS fields using publicly available \jwst/\NIRSpec\ PRISM spectroscopy.  
Our catalog comprises 3446 spectroscopically confirmed sources at 4 $\leq $ z < 14.2, including 3361 star-forming galaxies (SFGs), of which 539 are robust (S/N > 3) LAEs. 
We measure the fraction of LAEs with EW$_0$ > 25~\AA\ (X$_{\mathrm{Ly\alpha}}$) 
and trace its redshift evolution in two UV luminosity bins, namely -20.25 < M$_{\mathrm{UV}}$ < -18.75 and -21.75 < M$_{\mathrm{UV}}$ < -20.25. Within the fainter-UV range, X$_{\mathrm{Ly\alpha}}$ increases from z = 5 to z = 6 at 3$\sigma$ significance and subsequently declines toward higher redshifts with a significant monotonic trend at z > 6 and a 3$\sigma$ decrease between z = 6 and z = 12. 
In parallel, we investigate the physical properties of both LAEs and the full SFG population. We find that the stellar mass, UV slope $\beta$, stellar reddening, SFR, metallicity, sSFR, and burstiness (SFR$_{\mathrm{10Myr}}$/SFR$_{\mathrm{30Myr}}$) of LAEs remain approximately stable with redshift. The only exception is the mass-weighted age which decreases with increased redshift, as expected. Conversely, the properties of the full SFG population evolve significantly, progressively approaching the region of galaxy-property space occupied by LAEs as redshift increases. This suggests that galaxy evolution may enhance the intrinsic production and escape of \lya\ photons toward earlier epochs. We argue that this effect should be accounted when inferring the evolution of the neutral hydrogen content of the IGM from the observed visibility of \lya\ emission. To this end, we employ a physically motivated framework based on the Attenuation-Free Model, jointly accounting for galaxy evolution and IGM attenuation. Our observations favor reionization histories that begin early and proceed gradually over scenarios characterized by a rapid increase in the cosmic neutral hydrogen fraction.}


 \maketitle

\section{Introduction} \label{sec:intro}
The hydrogen Lyman-$\alpha$ (\lya) transition at a rest-frame wavelength of 1215.67~\AA\ is predominantly generated by recombination in ionized gas surrounding young, star-forming regions. Because \lya\ is a resonant transition, its propagation is strongly affected by neutral hydrogen in the interstellar medium (ISM), circumgalactic medium (CGM), and intergalactic medium (IGM) \citep{Dijkstra2017}. The observed line therefore encodes information about both the internal properties of galaxies and the ionization state of their large-scale environment.

Over the last twenty years, extensive spectroscopic studies of star-forming galaxies (SFGs) have established \lya\ as an important diagnostic of early galaxy evolution and of the neutral-hydrogen content of the IGM \citep[e.g.,][]{Fontana2010, Ono2010, Stark2010}. At fixed UV absolute magnitude, ground-based observations indicate that the fraction of \lya\ photons escaping from galaxies increases by approximately two orders of magnitude from the local Universe at z = 0 to the z = 6 Universe \citep[e.g.,][]{Konno2016}. These studies have also shown that galaxies with strong \lya\ emission differ systematically from the broader SFG population. In particular, \lya\ emitters (LAEs), commonly defined as galaxies with \lya\ rest-frame equivalent width EW$_0$ > 25~\AA\ \citep[e.g.,][]{Cassata2015, Ouchi_2020}, tend to have lower stellar masses, younger stellar populations, less dust attenuation, and bluer UV slopes $\beta$ than galaxies without detectable \lya\ emission (Non-LAEs) \citep[e.g.,][]{Gawiser2006, Marchi2019, Ouchi_2020, Napolitano2023, ChavezOrtiz2024, Iani2024}. 

Beyond the properties of galaxies, the incidence of \lya\ emission provides a sensitive probe of cosmic reionization. As the neutral fraction of the IGM increases toward earlier epochs, the transmission of \lya\ photons is expected to decrease \citep[e.g.,][]{Dijkstra2014}. During the Epoch of Reionization (EoR), this should produce a rapid decline in the fraction of SFGs displaying \lya\ emission above a given equivalent-width threshold, usually referred to as the \lya\ fraction or \lya\ visibility, X$_{\mathrm{Ly}\alpha}$. Measurements of the cumulative EW$_0$ distribution, typically integrated above thresholds of approximately 25--50~\AA, have therefore been translated into constraints on the evolution of the volume-averaged neutral-hydrogen fraction and on the timeline of reionization \citep[e.g.,][]{Mason2018a, Pentericci_2018b, jung2020}. Such inferences have generally assumed that the intrinsic \lya\ properties of SFGs evolve only weakly between the end of reionization and the populations observed at z $\sim$ 6.5, despite the $\sim$ 100--200~Myr of galaxy evolution separating these epochs.
\begin{figure*}[!ht]
\begin{minipage}{\textwidth}
\centering
\includegraphics[width=\linewidth]{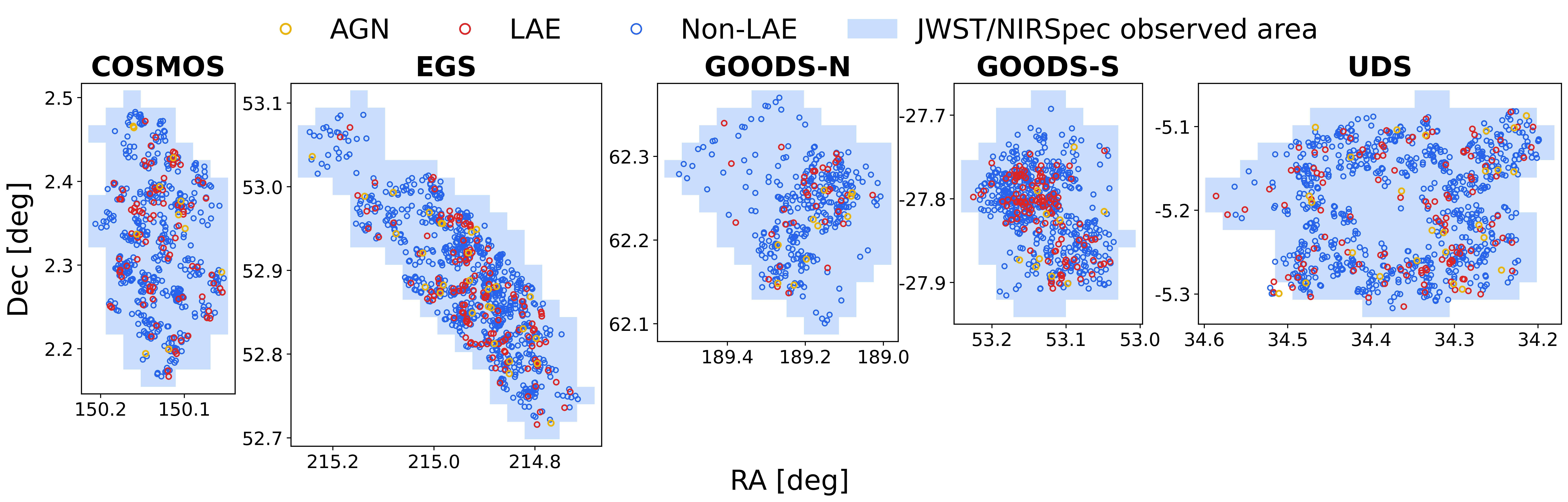}
\end{minipage}
\caption{Spatial distribution of the 3446 sources considered in this work across the five CANDELS fields. AGNs are shown as yellow circles, while SFGs are divided into LAEs and Non-LAEs, represented by red and blue circles, respectively. The pale-blue shaded regions indicate the areas covered by JWST/NIRSpec spectroscopy.}
\label{fig:Sky_distribution}
\end{figure*}

The advent of \jwst\ has fundamentally changed this observational picture \citep{Gardner2023}. Whereas previous studies were often based on only a few tens of \lya\ detections in each redshift interval, \jwst\ has now delivered spectra for hundreds of SFGs at z > 4, with secure redshifts enabled by multiple emission-line detections and continuum constraints near the \lya-break. 
These datasets have enabled statistically robust measurements of \lya\ visibility across several independent survey areas and new constraints on the evolution of the IGM neutral fraction \citep[e.g.,][]{Jones2024, Nakane2024, Napolitano2024, Tang2024B, Umeda2024, Witstok2024, Jones2025, Kageura2025, Umeda2025, Napolitano2026, Whitler2026, Chen2026}. Secure and tentative \lya\ detections have also extended this investigation into the cosmic-dawn regime at $z\simeq10$--13 \citep[i.e., GN-z11, JADES-GS-z13-1-LA, and EGS-z11-R0,][]{Bunker2023B, Witstok2025, Rodighiero2026}. These results have renewed discussion about whether the relatively weak evolution of X$_{\mathrm{Ly}\alpha}$ implies a slower increase in the neutral fraction, x$_{\mathrm{HI}}$, toward higher redshifts, or instead reflects an evolution in the intrinsic properties of the galaxy population \citep[e.g.,][]{Ferrara2024, Kageura2025}. One possibility is that typical SFGs at higher redshifts increasingly resemble lower-redshift LAEs in terms of their ISM properties and ionizing-photon production \citep[$\xi_{\rm ion}$, e.g.,][]{Simmonds2023, Begley2026}, thereby enhancing the intrinsic visibility of \lya\ even as the IGM becomes more neutral.

An additional complication is the strong spatial inhomogeneity of reionization \citep[e.g.,][]{Castellano2016, Keating2020}. Differences in the abundance and extent of ionized regions can produce substantial variations in \lya\ transmission among independent lines of sight, affecting measurements based on individual survey fields \citep[e.g.,][]{Jones2024, Napolitano2024, Napolitano2026}. In this work, we mitigate this source of cosmic variance by combining all publicly available \jwst/\NIRSpec\ spectra of SFGs in the five CANDELS fields. The compiled sample allows us to measure the average evolution of \lya\ visibility over multiple independent lines of sight. We interpret the resulting trend within the context of evolving ISM properties of the underlying SFG population, testing the main galaxy evolution interpretation of high-redshift galaxies progressively approaching the locus occupied by strong \lya\ emitters at lower redshifts.\\

This paper is organized as follows. Sect.~\ref{sec:Data_and_sample_selection} describes the spectroscopic dataset, while Sect.~\ref{sec:Method} presents the methods used to derive the relevant spectral and physical measurements. In Sect.~\ref{sec:XLya}, we measure the average \lya\ fraction, X$_{\mathrm{Ly}\alpha}$, and investigate its redshift evolution. Sect.~\ref{sec:properties} examines the role of evolving galaxy properties, Sect.~\ref{sec:AFM_pred} introduces a physically motivated framework for predicting the intrinsic \lya\ emission from galaxy evolution, and Sect.~\ref{sec:reionization} discusses the implications for the history of cosmic reionization and the ionization state of the IGM. Our main conclusions are summarized in Sect.~\ref{sec:Conclusion}.\\
Throughout this work, we assume a $\Lambda$CDM concordance cosmological model ($H_0 = 70$ \kmsmpc, $\Omega_M = 0.3$, and $\Omega_{\Lambda} = 0.7$). Magnitudes are expressed in the AB system \citep{Oke1983}, and all equivalent widths are reported in the rest frame.

\section{Data} \label{sec:Data_and_sample_selection}
To investigate and compare the properties of LAEs and Non-LAEs, we compiled the largest possible sample of galaxies with \jwst\ spectroscopic observations in the well studied Cosmic Assembly Near-infrared Deep
Extragalactic Legacy Survey \citep[CANDELS,][]{Grogin2011, Koekemoer2011} extragalactic fields; i.e., the Cosmic Evolution Survey Field \citep[COSMOS,][]{Scoville2007}, the Extended Groth Strip \citep[EGS,][]{Davis2007}, the Great Observatories Origins Deep Survey \citep[GOODS-N and GOODS-S,][]{Giavalisco2004, Dickinson2004}, and the Ultra Deep Survey \citep[UDS,][]{Lawrence2007} fields. We restricted the analysis to sources with secure redshifts at z > 4 (see Sect.~\ref{sec:zspec}), since the wavelength coverage of the \NIRSpec\ PRISM configuration allows \lya\ emission to be probed only above this redshift. \\ 
In total, we considered 3446 galaxies at 4 $\leq$ z $\leq$ 14.2: 560 are from the COSMOS, 955 from the EGS, 367 from the GOODS-N, 676 from the GOODS-S, and 888 from the UDS fields. Spectra come from multiple JWST NIRSpec surveys. The most represented one is the CAPERS survey, which accounts for $\sim$36\% of the total sample, while the rest of the spectra are obtained through the public DAWN JWST Archive \citep[DJA v4.4,][see Sect.~\ref{sec:otherJWST} for more details]{Heintz2024dja, deGraaff2025}.\\
Duplicate spectra of the same galaxy observed in multiple surveys were identified by searching for sources separated by less than 0.3 arcsec. These cases were then visually inspected using the publicly available DJA viewer to confirm whether they corresponded to real duplicates. When the same galaxy was observed more than once, we retained only the spectrum with the longest exposure time, as this choice provides the highest signal-to-noise ratio (S/N).\\
Fig.~\ref{fig:Sky_distribution} shows the spatial distribution of our sample in the five CANDELS fields. Among the total sample, 3361 are SFGs\footnote{Throughout this work, we use the term SFGs to denote all galaxies not classified as AGNs. No additional selection was applied to remove quiescent galaxies.}, while 85 are broad-line AGNs (see Sect.~\ref{sec:AGN}). For our aims, throughout the paper, SFGs are further divided into LAEs and Non-LAEs based on \lya\ measurement and upper limit (see Sect.~\ref{sec:LyAmodel}).

\begin{figure*}[!ht]
\begin{minipage}{0.5\textwidth}
\centering
\includegraphics[width=\linewidth]{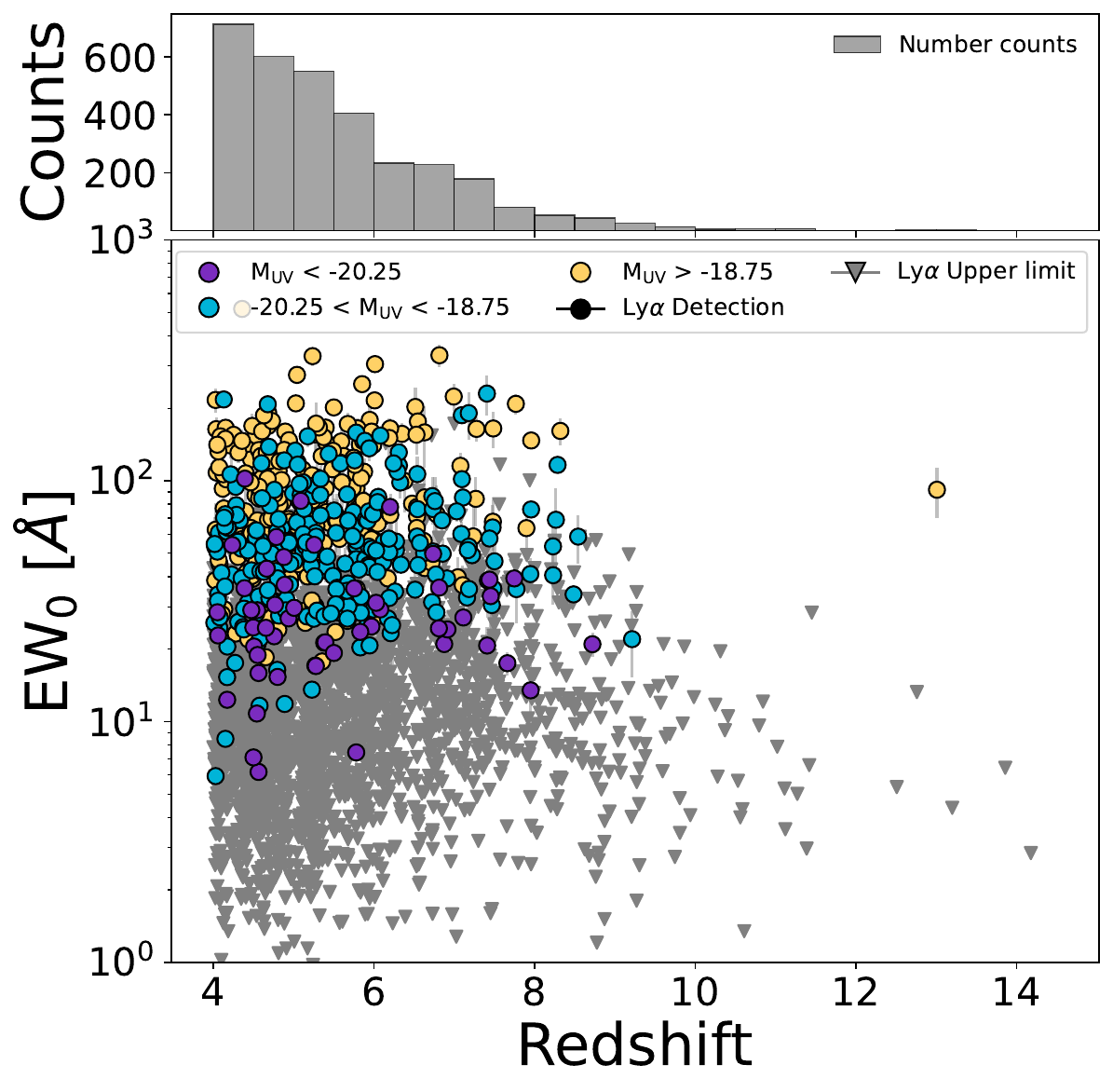}
\end{minipage}
\begin{minipage}{0.5\textwidth}
\centering
\includegraphics[width=\linewidth]{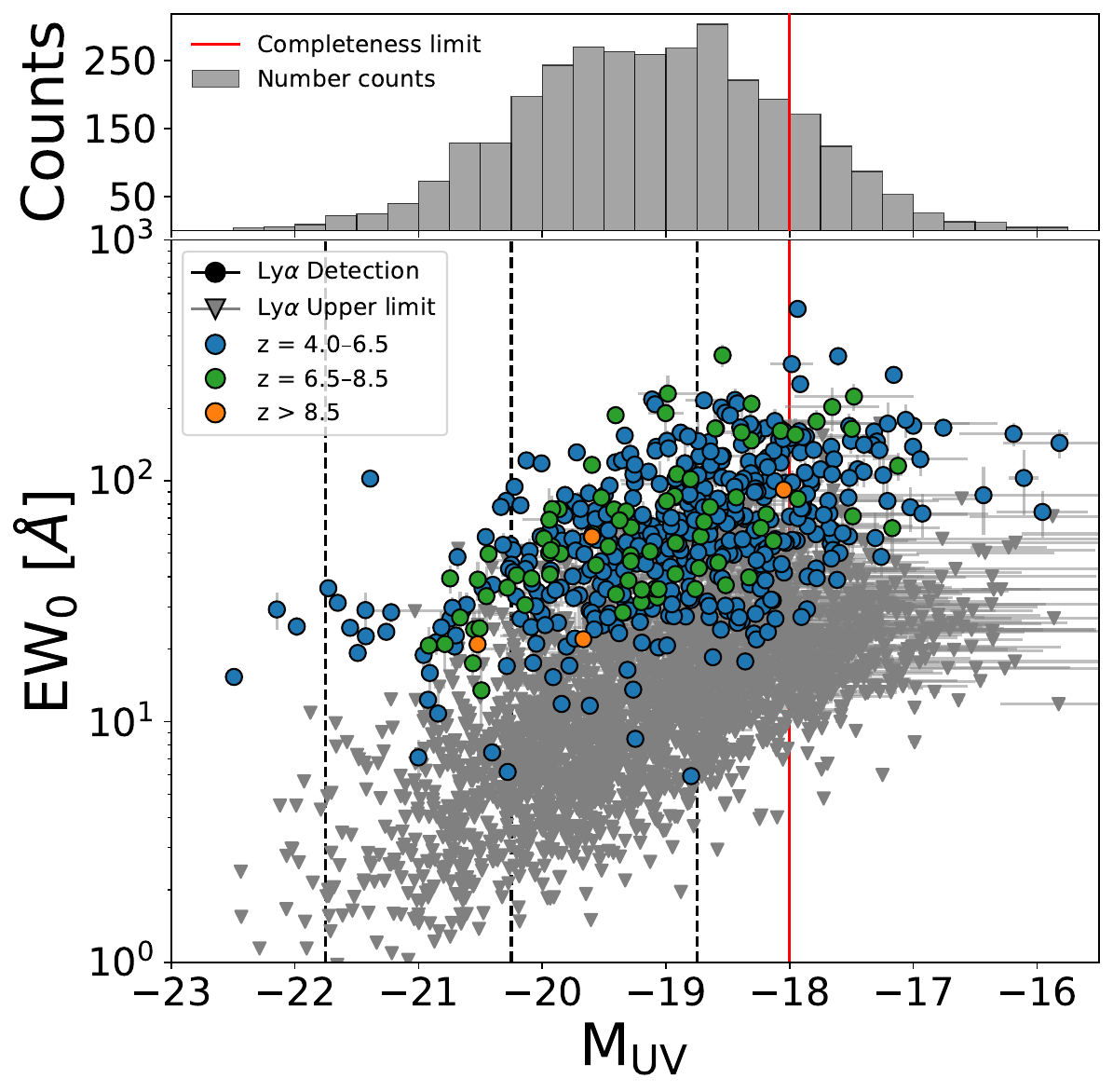}

\end{minipage}
\caption{Left: Rest-frame \lya\ equivalent width, EW$_0$, as a function of spectroscopic redshift. The 539 galaxies with robust \lya\ detection at S/N > 3 are shown as circles, while triangles indicate 1$\sigma$ upper limits EW$_{\mathrm{0,lim}}$ for the SFG population. Symbols are color-coded by M$_{\mathrm{UV}}$. Right: Distribution of \lya\ EW$_0$ as a function of M$_{\mathrm{UV}}$ at z > 4. The same symbols are used as in the left panel, while colors indicate spectroscopic redshift. The vertical dashed black lines at M$_{\mathrm{UV}}$ = -21.75, –20.25, and –18.75 delimit the adopted UV-luminosity intervals. The bright and intermediate regimes contain 16.3\% and 52.7\% of the SFGs at z > 4.5, respectively. The sample is complete down to a 60\% relative completeness limit of M$_{\mathrm{UV}}$ = -18, indicated by the solid red line.}
\label{fig:EW0compare}
\end{figure*}

\subsection{CAPERS data}\label{sec:CAPERS}
The spectroscopic data used in this work are part of the CANDELS-Area Prism Epoch of Reionization Survey (CAPERS; GO-6368, PI M. Dickinson), a JWST/NIRSpec PRISM/CLEAR program targeting galaxies in the CANDELS legacy fields. CAPERS is designed to cover 21 NIRSpec pointings, equally distributed among COSMOS, UDS, and EGS, with seven pointings per field. The PRISM/CLEAR setup provides continuous spectral coverage from approximately 0.6 to 5.3 $\mu$m at low spectral resolution, R $\sim$ 100. Each pointing consists of three independent micro-shutter assembly configurations, each with an integration time of about 5690 s. Depending on target priority and slit assignment, individual sources can therefore be observed in one, two, or three configurations, corresponding to total exposure times of approximately 1.6, 3.2, or 4.7 hr, respectively. The CAPERS pointings included in this work represent 94\% of the anticipated complete CAPERS observations.
A complete description of the target selection and survey strategy will be presented in the CAPERS survey paper.

The NIRSpec data were reduced with the STScI\footnote{\url{https://jwst-pipeline.readthedocs.io/en/latest/index.html}} JWST Calibration Pipeline version 1.20.2 \citep{Bushouse2025}, using Calibration Reference Data System mapping 1464. The reduction broadly followed the procedures described by \cite{ArrabalHaro2023}, with additional steps tailored to improve the quality of low-resolution NIRSpec spectra. In the detector-level processing, we used the \texttt{clean\_flicker\_noise} correction to reduce residual $1/f$ noise. During the \texttt{calwebb\_spec2} stage, we applied an updated flat-field correction. For slits containing more than one source, including primary targets and known or serendipitous companions, the 2D spectra were combined using an asymmetric nodding approach to limit contamination from neighboring objects \citep[e.g.,][]{Napolitano2025a}. The final rectified 2D spectra were produced after the \texttt{calwebb\_spec3} stage, and the corresponding 1D spectra were obtained through optimal extraction, following \cite{Horne1986}.

We also checked the absolute flux scale of the extracted spectra against NIRCam photometry from the publicly available ASTRODEEP catalog \citep{Merlin2024}. Synthetic fluxes were computed by integrating each spectrum through the F090W, F115W, F150W, F200W, F277W, F356W, and F444W filter transmission curves and compared with the observed broadband photometry. Only filters redward of the \lya-break and with synthetic photometry detected at S/N > 5 were used. For each source, we derived a single wavelength-independent multiplicative correction from the weighted mean of the individual filter ratios and applied it consistently to both the flux density and uncertainty arrays. We verified that the median correction factors measured in the individual filters show no significant monotonic dependence on observed wavelength through a non-parametric Spearman correlation test (p-value = 0.7), supporting the use of a single wavelength-independent correction for each source. The median multiplicative correction for the galaxy population is 1.4, with the 16th and 84th percentiles of the distribution corresponding to 1.1 and 2.0, respectively. This correction only changes the absolute flux normalization and therefore does not affect equivalent widths, UV continuum slopes ($\beta$), or physical parameters derived independently from the photometric SED fitting.

\subsection{CANDELS data from DJA archive} \label{sec:otherJWST}
To supplement the CAPERS sample, we retrieved all publicly available \jwst\ \NIRSpec\ PRISM observations in the CANDELS fields included in version 4.4 of the DJA archive. We selected only the most secure spectroscopic identifications, corresponding to \texttt{grade = 3} in the DJA catalog, and restricted the sample to sources at z > 4, consistently with the redshift range investigated in this work. We verified that the DJA and internal CAPERS reductions provide highly consistent spectra for the sources available in both data sets (see Sect.~\ref{sec:CAPERS_vs_DJA_app}). \\
This resulting DJA dataset includes: 783 spectra (i.e., $\sim$23\% of the full sample) from the JWST Advanced Deep Extragalactic Survey \citep[JADES; GTO-1180, 1181, 1210, 1212, 1286, 1287;][]{Bunker2020, Bunker2023, Eisenstein2023}; 678 spectra (i.e., $\sim$20\%) from the Red Unknowns: Bright Infrared Extragalactic Survey \citep[RUBIES GO-4233;][]{deGraaff2025}; 174 spectra (i.e., $\sim$5\%) from the Cosmic Evolution Early Release Science Survey \citep[CEERS ERS-1345, DDT-2750;][]{Arrabal_Haro2023Nature, Finkelstein2025}. An additional 580 spectra (i.e., $\sim$16\%) are from the combined GTO-1211, GTO-1213, GTO-1214, GTO-1215, GO-2198, GO-2565, GO-3215, GO-4106, GO-5224, GO-6585, and DDT-6541 programs. \\
To ensure a homogeneous absolute flux calibration across the full dataset, we applied to all PRISM spectra the same photometric rescaling procedure described for the CAPERS observations.

\section{Methods} \label{sec:Method}

\subsection{Spectroscopic redshifts} \label{sec:zspec}
To determine the spectroscopic redshifts, we used the redshifts reported in the available catalogs as initial solutions. For CAPERS sources, these redshifts were obtained through a combination of automated fitting and line-identification tools, including \textsc{bagpipes} \citep{Carnall2018}, \textsc{cigale} \citep{Boquien2019}, \textsc{LiMe} \citep{Fernandez2024}, \textsc{marz} \citep{Hinton2016}, and \texttt{msaexp} \citep{Brammer2022_msaexp}, followed by visual assessment from at least two independent members of the collaboration. The reliability of each solution was encoded through a quality flag, with values of 3 and 4 corresponding to the most secure identifications, supported by several detected emission lines \citep[for the adopted flag definition see][]{Pentericci2018}. A complete description of the CAPERS redshift-validation procedure will be presented in the survey overview paper. For JADES targets, we used the publicly released redshift catalogs of \cite{Curtis-Lake2026} and \cite{Scholtz2026}, retaining sources assigned to the secure A, B, or C confidence classes. For the remaining objects drawn from the DJA archive, the reference redshifts were the \texttt{msaexp}-based solutions classified as secure with a \texttt{grade = 3} \citep{Brammer2022_msaexp}.

Starting from these high-confidence catalog measurements, we selected galaxies at z > 4. We then inspected the 1D and 2D spectra of every candidate to independently assess the adopted redshift. This inspection relied on the position of the \lya\ continuum break, together with any available rest-frame UV and optical emission features. When one or more lines were detected at S/N > 3, we refined the redshift using their observed line centroids \citep[e.g.,][]{Napolitano2025a, Castellano2026}. 

In measuring the line centroids, we accounted for broadening introduced by the wavelength-dependent instrumental resolution. The width of the expected observed Gaussian profile was taken to be $\sigma_R (\lambda_{\mathrm{obs}}) [\AA] = \lambda_{\mathrm{obs}} / 2.355 R(\lambda_{\mathrm{obs}})$, where $R(\lambda_{\mathrm{obs}})$ was derived from the nominal \jwst/\NIRSpec\ resolution curve for a source uniformly illuminating the slit\footnote{\url{https://jwst-docs.stsci.edu/jwst-near-infrared-spectrograph/nirspec-instrumentation/nirspec-dispersers-and-filters\#gsc.tab=0}}. When multiple significant features were available, the final spectroscopic redshift was obtained from the uncertainty-weighted mean of the individual line-based measurements. The rest-frame UV and optical transitions used for this purpose were H$\alpha$, [OIII]$\lambda5007$, [OIII]$\lambda4959$, H$\beta$, and [OII]$\lambda\lambda3727,3729$, detected in 2813, 2822, 819, 2512, and 1373 sources, respectively.

For the remaining 36 cases (i.e., 1\% of the sample), no emission line from the above list was detected at S/N > 3. In these cases, we retained the reference redshift from the parent catalog after verifying that it was consistent with the position of a securely identified \lya-break. Nearly half of these cases lie at the current spectroscopic high-redshift frontier of z > 9, and all their adopted redshifts agree with values previously reported in the literature \citep[e.g.,][]{Curtis-Lake2023, Carniani2024, Hainline2024, Witstok2024, Kokorev2025b, Tang2025, Taylor2025, Rodighiero2026}. \\
Among all confirmed sources at z > 9, twelve galaxies have spectroscopic redshifts that have not been previously reported. are CAPERS sources and will be discussed in detail in Donnan et al. (in prep.), while the remaining two were identified in the GO-5224 MoM (Mirage or Miracle) survey\footnote{The two galaxies are MoM-UDS-131711 (RA = 34.280123, DEC = -5.160729) at z = 9.984 $\pm$ 0.004 and MoM-UDS-132380 (RA = 34.261754, DEC = -5.159822) at z = 9.542 $\pm$ 0.004.}.  
Including these new identifications, our final sample contains a total of 63 galaxies at z > 9 in the CANDELS fields.

The resulting spectroscopic redshift distribution is shown in the left panel of Fig.~\ref{fig:EW0compare}. All galaxies in the final \NIRSpec\ PRISM sample show a securely identified \lya\ break. Moreover, our independently validated redshifts remain consistent with the CAPERS, JADES, or DJA values used for the initial selection, with differences smaller than |$\Delta$z|=0.1 in all cases.

\subsection{AGN identification} \label{sec:AGN}
Recent studies have shown that broad-line AGN, including both little red dots (LRDs) and little blue dots (LBDs), exhibit strong \lya\ emission \citep[e.g.,][]{Geris2026, Morishita2026, Jones2026, Torralba2026}. This enhancement is likely driven by AGN photoionization and emission from the broad-line region (BLR), rather than exclusively by star formation.

Direct evidence for this interpretation has been obtained from higher-resolution \NIRSpec\ spectroscopy. For example, \cite{Tang2026} resolved both narrow and broad \lya\ components in A2744-QSO1 at $z\sim7$, finding that the broad component dominates the total line flux. 
Similarly, \cite{Geris2026} identified a broad \lya\ component in stacked LRD spectra with approximately twice the flux of the narrow component and a profile comparable to that of broad H$\alpha$, again supporting a BLR origin. Their LRD and LBD stacks also display \lya\ EW$_0$ values approximately four and two times larger, respectively, than those of star-forming galaxies selected at similar M$_{\mathrm{UV}}$. Consequently, including broad-line AGNs (BLAGNs) would preferentially add systems whose measured \lya\ EW$_0$ is boosted and cannot be reliably separated from the narrow star-forming component in PRISM spectra. We therefore exclude all BLAGNs from the star-forming galaxy sample used to investigate \lya\ visibility and its dependence on galaxy properties.

To identify BLAGN within our sample, we first visually inspected the \NIRSpec\ PRISM spectra and selected sources showing possible broad Balmer emission. These candidates were subsequently examined with a quantitative profile-fitting procedure based on the approach adopted in \cite{Napolitano2026} \citep[see also,][]{Taylor2024, Juodzbalis2025}. We modeled the H$\alpha$ emission whenever this line was available. Otherwise, we fitted the full H$\beta$+[OIII]$\lambda\lambda$4959,5007 line complex. For each narrow component, the centroid was fixed to the wavelength expected from the spectroscopic redshift, while its width $\sigma_R$ was set by the wavelength-dependent \NIRSpec\ PRISM resolution. For each source, we compared two alternative models. The first model consisted of the narrow Balmer component, together with the [OIII] doublet when fitting H$\beta$, superimposed on a linear continuum. The second model included the same narrow-line components and continuum, with the addition of a broad Gaussian associated with the Balmer line. The continuum intercept and slope were free in both models. The broad-line observed FWHM was restricted to the range permitted by the instrumental resolution, with an upper bound of 15,000~km~s$^{-1}$. The posterior distributions were sampled with the \textsc{emcee} Markov chain Monte Carlo (MCMC) sampler \citep{Foreman_Mackey2013}. We used 32 walkers for the narrow-only model and 64 walkers for the model containing the additional broad component. Each chain was evolved for 800 burn-in steps, which were discarded, followed by 2,500 production steps. Parameter estimates were obtained from the posterior medians, with uncertainties corresponding to the 16th and 84th percentiles. We required the best-fit broad FWHM already de-convolved by the resolution of the instrument to exceed 1,000~km~s$^{-1}$. Finally, the integrated broad-to-narrow Balmer flux ratio was required to be at least 10\%. This additional condition reduces the probability that fluctuations in the spectral noise are interpreted as a weak broad component. Applying the described criteria, we identified 85 sources as BLAGNs, with FWHM values ranging from 1,030--10,800~km~s$^{-1}$.

We note that this procedure is not designed to identify narrow-line AGNs (NLAGNs). A robust NLAGN classification generally requires several securely detected rest-frame UV and optical emission lines, together with metallicity, ionization parameter, and gas density measurements, in order to robustly compare to AGN and stellar photoionization models \citep[e.g.,][]{Feltre2016, Gutkin2016, Nakajima2022} on diagnostic diagrams \citep[e.g.,][]{Hirschmann2019, Hirschmann2023, Mazzolari2024a}. In \cite{Tripodi2026}, we found no evidence for a significant NLAGN contribution in the stacked spectrum of \lya\ emitters at z > 4. Because that analysis was performed on a subsample of 287 galaxies considered here, it suggests that residual NLAGN contamination is very limited, although it cannot be robustly ruled out for individual objects. We therefore do not attempt a separate NLAGN selection in this work and use the term AGN hereafter exclusively for the 85 sources identified through their broad Balmer emission.

\subsection{M$_{\mathrm{UV}}$ and $\beta$ } \label{sec:Muv}
We measured the rest-frame UV continuum following the approach adopted in \cite{Napolitano2025a}. The UV slope was parameterized as $f_{\lambda}\propto\lambda^{\beta}$ and fitted over the rest-frame interval 1350--2600~\AA. Wavelengths below 1350~\AA\ were excluded to minimize possible contamination from the \lya\ damping wing \citep[e.g.,][]{Dottorini2025, Jecmen2026}, while the CIV$\lambda\lambda1548,1551$ and CIII]$\lambda1909$ regions were masked using widths set by the wavelength-dependent \NIRSpec\ resolution, $\sigma_R$. The continuum was fitted with \textsc{emcee} using 30 walkers, 10,000 burn-in steps, 50,000 production steps, and adopting a uniform prior of $-4<\beta<0$.
For each source, the reported UV slope corresponds to the median of the marginalized posterior distribution, while its uncertainty was estimated from the posterior standard deviation. All fits were additionally visually inspected to identify unreliable continuum solutions. We retained reliable $\beta$ values for 2653 galaxies (i.e., 77\% of the full sample). Unsuccessful fits were associated with incomplete continuum coverage, including the \NIRSpec\ detector gap, or with poorly constrained and non-Gaussian posteriors corresponding to S/N < 3 measurements.

For the 2653 galaxies with a reliable continuum fit, we derived the absolute UV magnitude, M$_{\mathrm{UV}}$, from the best-fitting \textsc{emcee} model by averaging the rest-frame flux density between 1450 and 1550~\AA. For an additional 607 sources, representing 18\% of the sample, M$_{\mathrm{UV}}$ was instead estimated from the available photometry. In these cases, we considered only filters whose transmission curves\footnote{\url{https://jwst-docs.stsci.edu/jwst-near-infrared-camera/nircam-instrumentation/nircam-filters\#gsc.tab=0}} include rest-frame 1500~\AA\ at the source redshift. In total, M$_{\mathrm{UV}}$ is therefore available for 3260 galaxies. The remaining 186 sources (5.4\% of the sample) lack a reliable spectroscopic or photometric constraint at 1500~\AA\ and were excluded from the subsequent analysis. The observed 60\% relative completeness limit of the sample is M$_{\mathrm{UV}}$ = -18, defined as the UV absolute magnitude value at which the number of detections decreases to 60\% of the peak of the observed distribution. In Sect.~\ref{sec:Muv_completeness_app}, we further investigate the redshift dependence of the UV absolute magnitude completeness limit. The resulting M$_{\mathrm{UV}}$ distribution is presented in the right panel of Fig.~\ref{fig:EW0compare}. 

The spectroscopic M$_{\mathrm{UV}}$ values were calculated after applying the flux-rescaling correction described in Sect.~\ref{sec:CAPERS}. Since the spectro-photometric correction modifies only the overall normalization of the spectrum, it does not affect the derived UV slope.

\subsection{\lya\ emission line measurements} \label{sec:LyAmodel}
We measured the \lya\ emission using the forward-modeling procedure developed in \cite{Napolitano2024} and subsequently applied in \cite{Napolitano2026}, which is designed for the low spectral resolution of \NIRSpec\ PRISM. 

The intrinsic \lya\ emission was modeled as a Gaussian with FWHM uniformly distributed between 100 and 1550~km~s$^{-1}$. A preliminary line flux was obtained through continuum-subtracted integration over five spectral pixels centered on the observed emission peak, and the Gaussian amplitude was assigned a uniform prior corresponding to 0.05--20 times this initial flux estimate. To reproduce the sharp continuum discontinuity across \lya, the emission profile was added to a two-level continuum model. Redward of \lya, the continuum followed a linear fit extending from three pixels beyond the line peak to 1900~\AA\ in the rest frame, while blueward of the line it was fixed to the median observed flux to account empirically for IGM absorption.
The complete model was convolved with a Gaussian kernel of width $\sigma_R(\lambda_{\mathrm{obs}})$ to reproduce the wavelength-dependent instrumental resolution. Posterior distributions were sampled with \textsc{emcee} using 10 walkers, 1,000 burn-in steps, and 20,000 production steps. We adopted the posterior medians as the \lya\ flux and EW$_0$ measurements, with uncertainties given by the 16th and 84th percentiles. All fitted lines and surrounding continua were visually inspected. The performance of this modeling approach was previously tested on 73 galaxies that are also included in the present sample, showing no systematic residuals \citep{Napolitano2026}.

Adopting a detection threshold of S/N > 3 on the integrated \lya\ flux, we identify 539 robust \lya\ emitters among the SFGs sample in the CANDELS fields at z > 4. Their distribution in redshift, M$_{\mathrm{UV}}$, and EW$_0$ are shown in Fig.~\ref{fig:EW0compare}. In terms of the absolute number of detections, this sample is approximately 3--15 times larger than those assembled in recent studies. In particular, \cite{Jones2025}, \cite{Napolitano2026}, \cite{Kageura2025}, \cite{Chen2026}, and \cite{Tang2024B} identified 150, 73, 60, 36, and 33 \lya\ emitters within parent samples of 795, 651, 586, 292, and 210 star-forming galaxies, respectively. The substantially larger number of detections obtained here primarily results from our analysis of all publicly available \jwst\ spectra of star-forming galaxies in the CANDELS fields, comprising 3446 unique sources inspected. Our catalog naturally includes several \lya\ emitters previously reported in the literature. For sources in common, we compared our measurements with published values from the largest data samples available to date \citep[i.e.,][]{Jones2025, Napolitano2026}. We found no significant systematic offset from the one-to-one relation. Throughout this work, however, we adopt only the measurements derived here to ensure that a uniform fitting procedure and consistent selection criteria are applied across all fields and redshift intervals.

To quantify the sensitivity to \lya\ emission independently of whether a line was detected, we also calculated a limiting rest-frame equivalent width, EW$_{0,\mathrm{lim}}$, for every galaxy. Following Equation~2 of \cite{Jones2024}, the limit was evaluated as
\begin{equation} \label{eq:EW0lim}
EW_{\mathrm{0,lim}} = \frac{\sqrt{2 \pi} \ E(\lambda ^{\lya}) \sigma_R (\lambda_{\mathrm{obs}})}{(1+z)F_{\lambda}^{\mathrm{cont\ }}}.
\end{equation} 
where $E(\lambda^{\lya})$ is the flux-density uncertainty at the expected \lya\ wavelength, $F_{\lambda}^{\mathrm{cont}}$ is the continuum flux density, and $\sigma_R(\lambda_{\mathrm{obs}})$ accounts for the instrumental broadening. These limits are used in Sect.~\ref{sec:XLya} to define the completeness of the \lya\ sample. The right panel of Fig.~\ref{fig:EW0compare} presents both the measured EW$_0$ values and the corresponding $1\sigma$ limits as functions of redshift and M$_{\mathrm{UV}}$.

\subsection{SED fitting} \label{sec:SED_fitting}
We derived the stellar and ISM properties of the 3361 SFGs in our final sample by revisiting the SED-fitting analysis presented by \cite{Llerena2026}. The quantities inferred from the fits include stellar mass, star formation rate (SFR), stellar reddening (E(B-V)), stellar metallicity (Z), and mass weighted age (age).

Briefly, the modeling was performed with version 1.2.0 of \textsc{bagpipes} \citep{Carnall2018}, adopting the stellar population synthesis models of \cite{Bruzual2003} assuming a \cite{Kroupa2001} initial mass function. For each galaxy, the redshift was fixed to the secure spectroscopic value. We fitted the multi-band photometric catalog of \cite{Merlin2024}, which combines \jwst/\NIRCam\ observations in F090W, F115W, F150W, F200W, F277W, F356W, F410M, and F444W with archival \hst\ imaging in F435W, F606W, F814W, F105W, F125W, F140W, and F160W. The precise wavelength coverage varies among the CANDELS fields: the EGS photometry does not include F435W, F090W, or F140W, whereas F105W is unavailable in the UDS field. We considered star formation histories using the non-parametric prescription of \cite{Leja2019}. The SFR was assumed to remain constant within eight independent age intervals. The four most recent bins cover lookback times of 0--3, 3--10, 10--30, and 30--100 Myr. The remaining four bins were spaced logarithmically between 100 Myr and t$_{\max}$=t$_{\mathrm{Universe}}$(z)$-$t$_{\mathrm{Universe}}$(z=20), thereby limiting the onset of star formation to z $\leq$ 20. This non-parametric approach yields average SFRs on short time-scales (10-20~Myr) consistent with those derived from Balmer dust-corrected luminosities \citep[e.g.,][]{Llerena2026, Santini2026}. The stellar metallicity was treated as a free parameter, with an allowed range from 0.01 to 0.5 Z$_{\odot}$ \citep{Llerena2026}. Dust attenuation was modeled using the \cite{Calzetti2000} law, allowing A$_{\mathrm{V}}$ to vary between 0 and 2 mag. Nebular emission was included self-consistently, assuming the nebular E(B-V) to be equal to the stellar reddening. Finally, the ionization parameter was allowed to vary freely over the interval -4 $\leq\log$U $\leq$ 0.

\section{\lya\ constraints on galaxy and IGM evolution in the CANDELS fields}\label{sec:lya_constraints}
\begin{figure*}[!ht]
\begin{minipage}{0.5\textwidth}
\centering
\includegraphics[width=\linewidth]{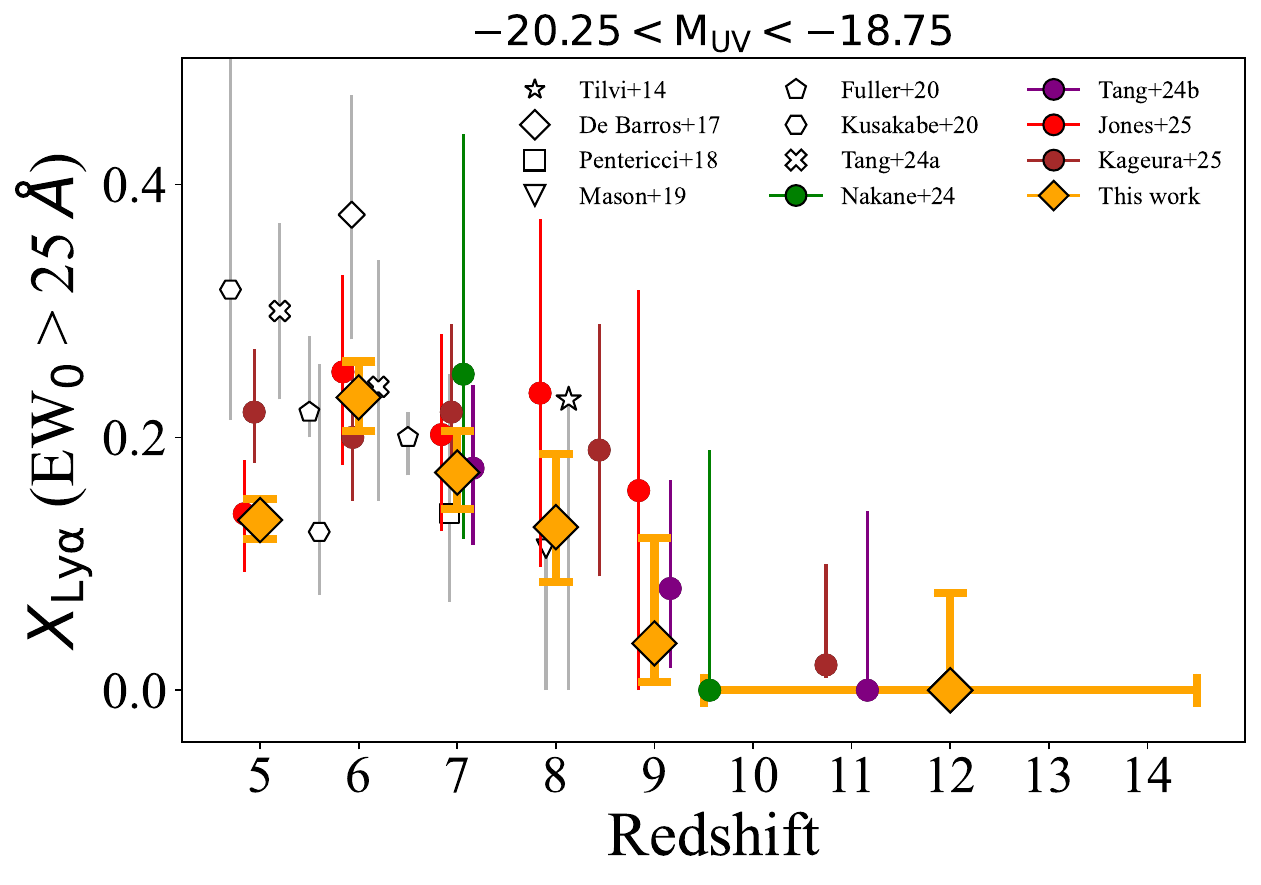}
\end{minipage}
\begin{minipage}{0.5\textwidth}
\centering
\includegraphics[width=\linewidth]{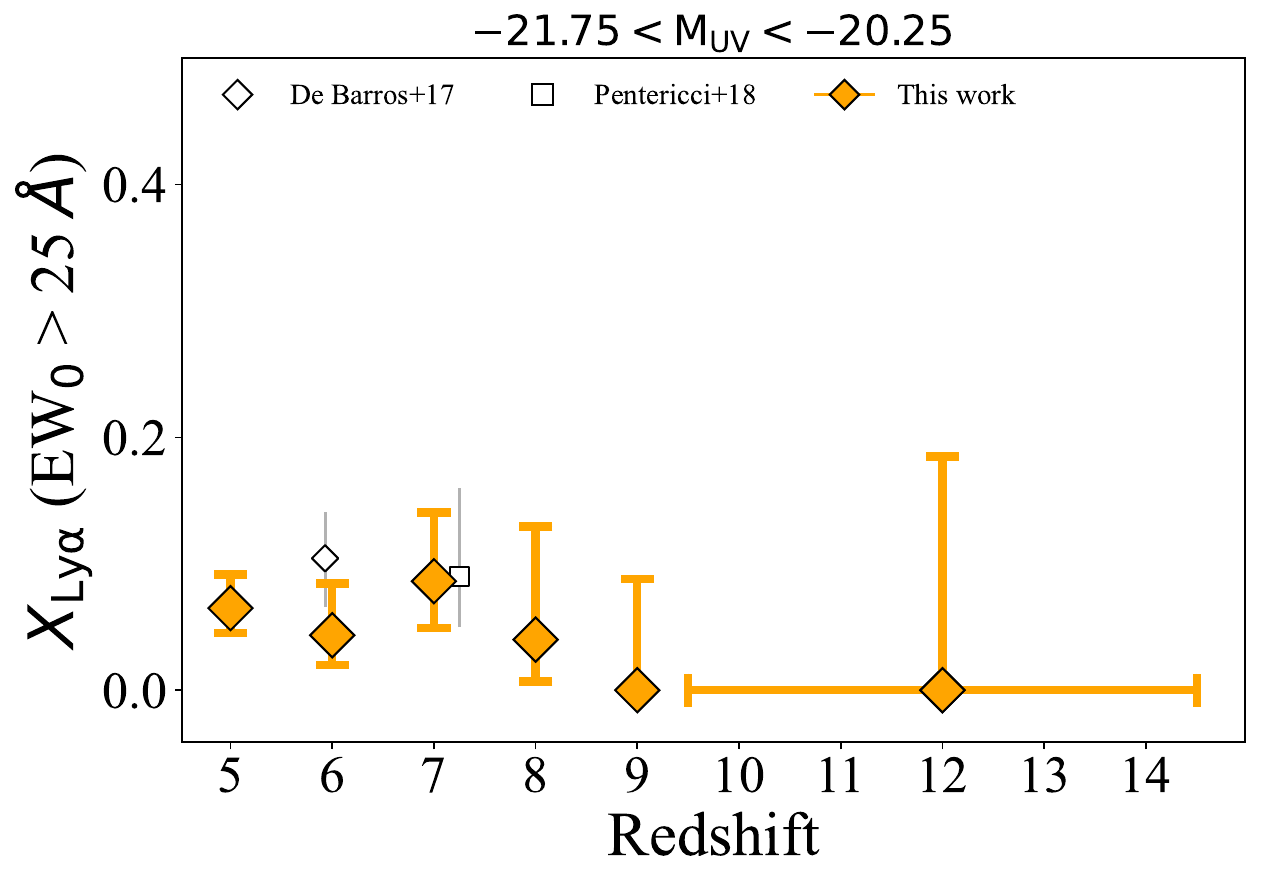}

\end{minipage}
\caption{Fraction of galaxies exhibiting rest-frame \lya\ EW$_0$> 25~\AA\ as a function of redshift within the UV absolute magnitude ranges -20.25 < M$_{\mathrm{UV}}$ < -18.75 (left) and -21.75 < M$_{\mathrm{UV}}$ < -20.25 (right). All reported measurements are averaged across multiple survey fields. Filled colored circles indicate previous \jwst-based results \citep{Nakane2024, Tang2024B, Jones2025, Kageura2025}, while the results of this work are shown as orange diamonds. Open symbols represent results from ground-based observations \citep{Tilvi2014,  DeBarros2017, Pentericci_2018b, Mason2019, Fuller2020, Kusakabe2020, Tang2024}. Uncertainties were derived using the binomial confidence intervals of \cite{Gehrels1986}. The measurements were obtained in redshift bins of width $\Delta$z = 1, except for the highest-redshift point, which combines all spectroscopically confirmed galaxies at 9.5 < z < 14.5. Points have been slightly offset along the redshift axis for visual clarity.}
\label{fig:Xlya_vs_z}
\end{figure*}
In this section, we extend the analysis of LAEs and Non-LAEs to all five CANDELS fields using \NIRSpec\ PRISM spectroscopy of SFGs at z > 4. We first present a homogeneous measurement of the \lya\ emitter fraction and investigate its gradual decline with increasing redshift across the EoR. We then interpret this trend by considering the combined effects of intrinsic changes in the galaxy population and in the ionization state of the IGM. In particular, we examine how the measured ISM properties of SFGs vary with redshift and compare the observed trends with model predictions.

\subsection{The observed evolution of the \lya\ fraction}\label{sec:XLya}
We investigated the redshift evolution of the \lya\ emitter fraction, X$_{\mathrm{Ly\alpha}}$, following the description adopted in \cite{Napolitano2026}. Briefly, X$_{\mathrm{Ly\alpha}}$ was calculated as the number of galaxies with a robust \lya\ detection at S/N > 3 and EW$_0$ > 25~\AA, divided by the total number of SFGs satisfying the selection criteria described below. Sources showing evidence of AGN activity were excluded, as described in Sect.~\ref{sec:AGN}. For the X$_{\mathrm{Ly\alpha}}$ measurement, we further restricted the sample to the intermediate and bright absolute UV magnitude ranges of -20.25 < M$_{\mathrm{UV}}$ < -18.75 and -21.75 < M$_{\mathrm{UV}}$ < -20.25, respectively. The intermediate range is commonly used in previous literature \citep[e.g.,][]{Stark2011, Jones2024, Kageura2025}, facilitating a direct comparison. We note that the defined intermediate UV range contains 52.7\% of the sample and lies well above our estimated 60\% relative completeness limit of M$_{\mathrm{UV}}$ = -18 (see Sect.~\ref{sec:Muv}). By comparison, only 16.3\% of the sample occupies the brighter interval. We do not extend the primary analysis to substantially fainter magnitudes explored by \cite{Kageura2025} (i.e., -18.75 < M$_{\mathrm{UV}}$ < -17.25) because the five considered unlensed CANDELS fields become affected by incompleteness at these magnitudes (see Sect.~\ref{sec:Muv_completeness_app}). We further note that all galaxies within the reported completeness limit have a secure UV continuum detection.
Controlling the UV-luminosity range is also necessary because \lya\ EW$_0$ depends strongly on M$_{\mathrm{UV}}$ \citep[Fig.~\ref{fig:EW0compare}, see also][]{Nakane2024, Napolitano2024, Jones2025, Gagnon-Hartman2026}. As discussed in Sect.~\ref{sec:properties}, this relation is partly driven by the intrinsic properties of galaxies, which regulate the production and escape of \lya\ photons through the ISM. Selection effects also contribute at the faint end, where the increasing EW$_{0,\mathrm{lim}}$ means that only the strongest lines can be recovered (Fig.~\ref{fig:EW0compare}). In contrast, the absence of sources with EW$_0$ > 100~\AA\ at the bright end cannot be attributed to limited sensitivity, since the corresponding EW$_{0,\mathrm{lim}}$ values are at least an order of magnitude smaller, but is instead physically motivated.

We imposed an additional sensitivity criterion on the \lya\ emission to ensure that the numerator and denominator of X$_{\mathrm{Ly\alpha}}$ were defined over a uniformly complete sample. Only galaxies with a 
limiting rest-frame equivalent width of EW$_{0,\mathrm{lim}}$ < 25~\AA\ were retained. Sources with shallower limits were removed from the statistical sample, irrespective of whether an emission feature was present. This selection guaranties that included spectra were sufficiently sensitive to detect \lya\ above the adopted EW$_0$ threshold, which is set by the source continuum and redshift (see Eq.~\ref{eq:EW0lim}): it avoids both an upward bias caused by detecting only bright lines in shallow spectra and a downward bias from including continuum-faint galaxies for which a line with EW$_0$ $\simeq$ 25~\AA\ would remain undetectable. After applying all criteria, the final sample contains 1084 (335) SFGs, of which 173 (19) contribute to the numerator of X$_{\mathrm{Ly\alpha}}$ in the intermediate (bright) UV range. Throughout the analysis in this section, X$_{\mathrm{Ly\alpha}}$ was computed using the observed EW$_0$ values, without applying any further \lya\ slit-loss corrections \citep[e.g.,][]{Tang2024}.

We measured X$_{\mathrm{Ly\alpha}}$ in six redshift intervals centered at z = 5, 6, 7, 8, 9, and 12. The first five bins have a width of $\Delta$z = 1, while the highest-redshift bin combines all galaxies at 9.5 < z < 14.5. Confidence intervals were derived from binomial statistics following \cite{Gehrels1986}.
The resulting evolution of the \lya\ emitter fraction averaged over the five CANDELS fields is reported in Fig.~\ref{fig:Xlya_vs_z} and in Table~\ref{tab:Xlya}. In the intermediate UV range, we observe a 3.1$\sigma$ increase in X$_{\mathrm{Ly\alpha}}$ between z = 5 and z = 6, followed by a gradual decrease from z = 6 toward higher redshifts. Overall, the difference between the z = 6 bin and the z = 12 measurement corresponds to a 3$\sigma$ decrease. 
To further test the observed decline in \lya\ visibility at z > 6, we performed a non-parametric Spearman correlation test, which confirms that the redshift evolution is significant (p-value = 0.02). In the bright UV range, the number of detected LAEs remains $\leq$10 in each redshift bin. Although X$_{\mathrm{Ly\alpha}}$ shows a tentative decline at z > 7, the limited number of emitters results in large uncertainties and prevents a statistically significant detection of redshift evolution ($\rho$ = -0.9, p-value = 0.1).
\begin{table}
\caption{Observed fraction of \lya-emitting galaxies with EW$_0$ > 25~\AA\ in the intermediate and bright UV ranges.}
    \label{tab:Xlya}
    \centering
    \begin{tabular}{ccc}
    \hline \noalign{\smallskip}
         & \multicolumn{2}{c}{X$_{\mathrm{Ly\alpha}}$ (EW$_0$ > 25 \AA) in CANDELS fields} \\
        \noalign{\smallskip}
        Redshift & -20.25 < M$_{\mathrm{UV}}$ < -18.75 & -21.75 < M$_{\mathrm{UV}}$ < -20.25 \\
        \hline \noalign{\smallskip}
        5 & $13.5_{-1.5}^{+1.7} \ \% \ \mathrm{(i.e., 70/520)}$ & $6.5_{-2.0}^{+2.6} \ \% \ \mathrm{(i.e., 10/154)}$ \\ [3pt]
        6 & $23.2_{-2.6}^{+2.8} \ \% \ \mathrm{(i.e., 63/272)}$ & $4.3_{-2.3}^{+4.1} \ \% \ \mathrm{(i.e., 3/69)}$ \\ [3pt]
        7 & $17.2_{-2.9}^{+3.3} \ \% \ \mathrm{(i.e., 31/180)}$ & $8.6_{-3.7}^{+5.4} \ \% \ \mathrm{(i.e., 5/58)}$ \\ [3pt]
        8 & $12.9_{-4.3}^{+5.8} \ \% \ \mathrm{(i.e., 8/62)}$ & $4.0_{-3.3}^{+8.9} \ \% \ \mathrm{(i.e., 1/25)}$ \\ [3pt]
        9 & $3.7_{-3.1}^{+8.3} \ \% \ \mathrm{(i.e., 1/27)}$ & $0_{-0}^{+8.8} \ \% \ \mathrm{(i.e., 0/20)}$ \\ [3pt]
        12 & $0_{-0}^{+7.7} \ \% \ \mathrm{(i.e., 0/23)}$ & $0_{-0}^{+18} \ \% \ \mathrm{(i.e., 0/9)}$ \\ [3pt]
        \hline
    \end{tabular}
\end{table}

Within the intermediate UV range, the increase in X$_{\mathrm{Ly\alpha}}$ from z = 5 to z = 6, corresponding to the post-reionization epoch and its late stages \citep[e.g.,][]{bosman22}, is consistent with a similar trend reported in ground-based results. These studies showed that the escape of \lya\ photons from galaxies becomes progressively more efficient toward higher redshift, increasing by approximately two orders of magnitude from z = 0 to z = 6, as a consequence of the evolving properties of the galaxy population \citep[e.g.,][]{Konno2016}. However, at higher redshifts, between z = 6 and z = 12, our measurements confirm a robust gradual monotonic decline in X$_{\mathrm{Ly\alpha}}$ with increasing redshift, associated with the increasing neutral hydrogen fraction (x$_{\mathrm{HI}}$) of the IGM. Such a gradual evolution contrasts with the more abrupt decrease between z $\simeq$ 6 and 7 inferred from pre-\jwst\ observations, which was widely interpreted as evidence for a rapid evolution of the cosmic reionization history \citep[e.g.,][]{Stark2010, Mason2018a, Pentericci_2018b}. This interpretation generally relied on the assumption that the intrinsic \lya\ EW$_0$ distribution of SFGs evolved only weakly between the post-reionization population and galaxies at z $\simeq$ 7. 

A direct comparison between ground-based and \jwst-based measurements of the EW$_0$ distribution and X$_{\mathrm{Ly\alpha}}$ is complicated by different observational selection effects. 
In particular, ground-based measurements at z = 6--7 were obtained from samples constructed using heterogeneous photometric selections and for which \lya\ provided the primary means of spectroscopic redshift confirmation. This complicates a direct comparison of their EW$_0$ distributions with those measured by \jwst, as the \jwst/NIRSpec observations of SFGs at z = 6--7 do not rely on selections based on the presence of the \lya\ line. Conversely, \jwst/\NIRSpec\ MOS observations may underestimate the total \lya\ flux because resonant scattering produces spatially extended emission that is not fully captured by the narrow \NIRSpec\ slit apertures. Estimated slit losses of approximately 20--30\% could therefore bias \jwst-based measurements of X$_{\mathrm{Ly\alpha}}$ toward lower values \citep[e.g.,][]{Nakane2024, Tang2024, Bhagwat2025, Napolitano2026}. For these reasons, we focus here on the redshift evolution of X$_{\mathrm{Ly}\alpha}$ measured exclusively with \jwst, comparing redshift bins constructed using homogeneous target-selection cuts, observational setup, and the same analysis.

\begin{figure*}[ht!]
    \centering

    \begin{subfigure}[b]{0.33\textwidth}
        \includegraphics[width=\linewidth]{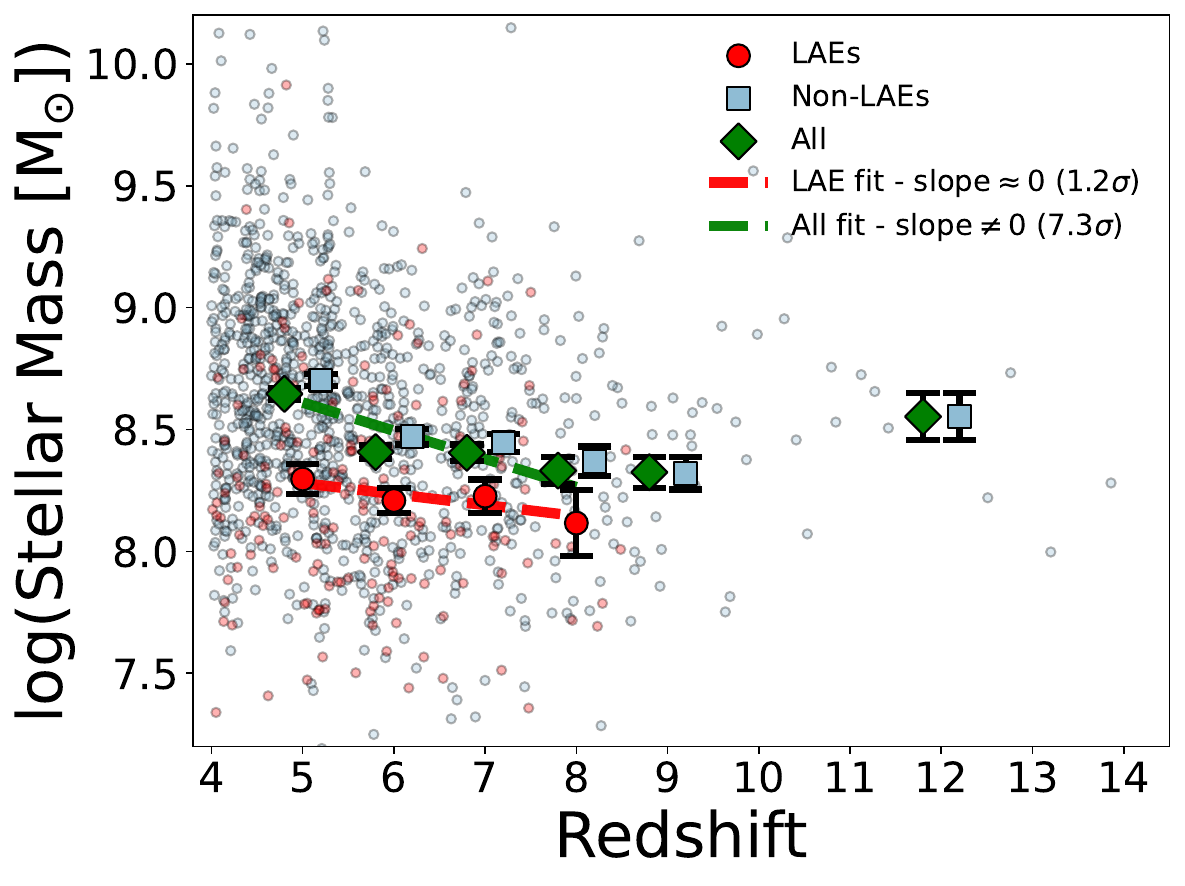}
    \end{subfigure}
    \begin{subfigure}[b]{0.33\textwidth}
        \includegraphics[width=\linewidth]{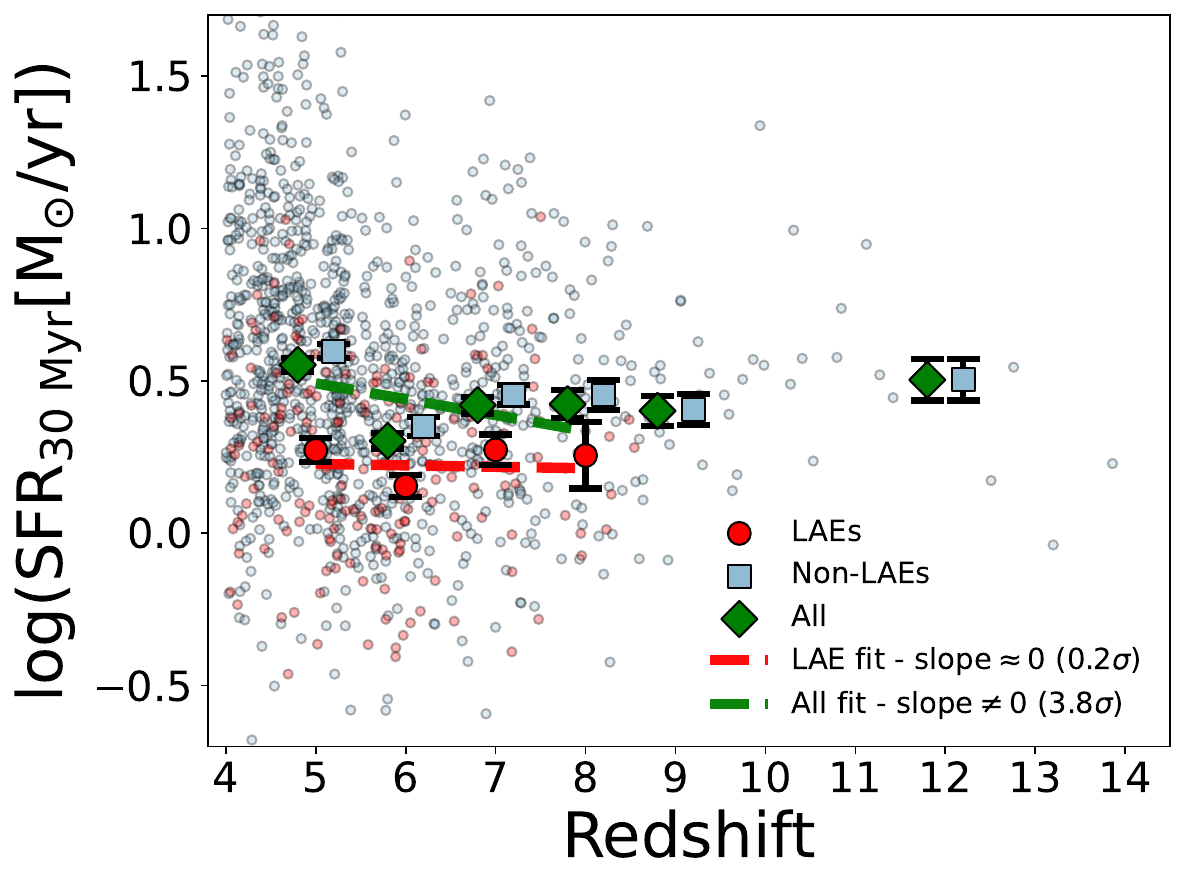}
    \end{subfigure}
    \begin{subfigure}[b]{0.33\textwidth}
        \includegraphics[width=\linewidth]{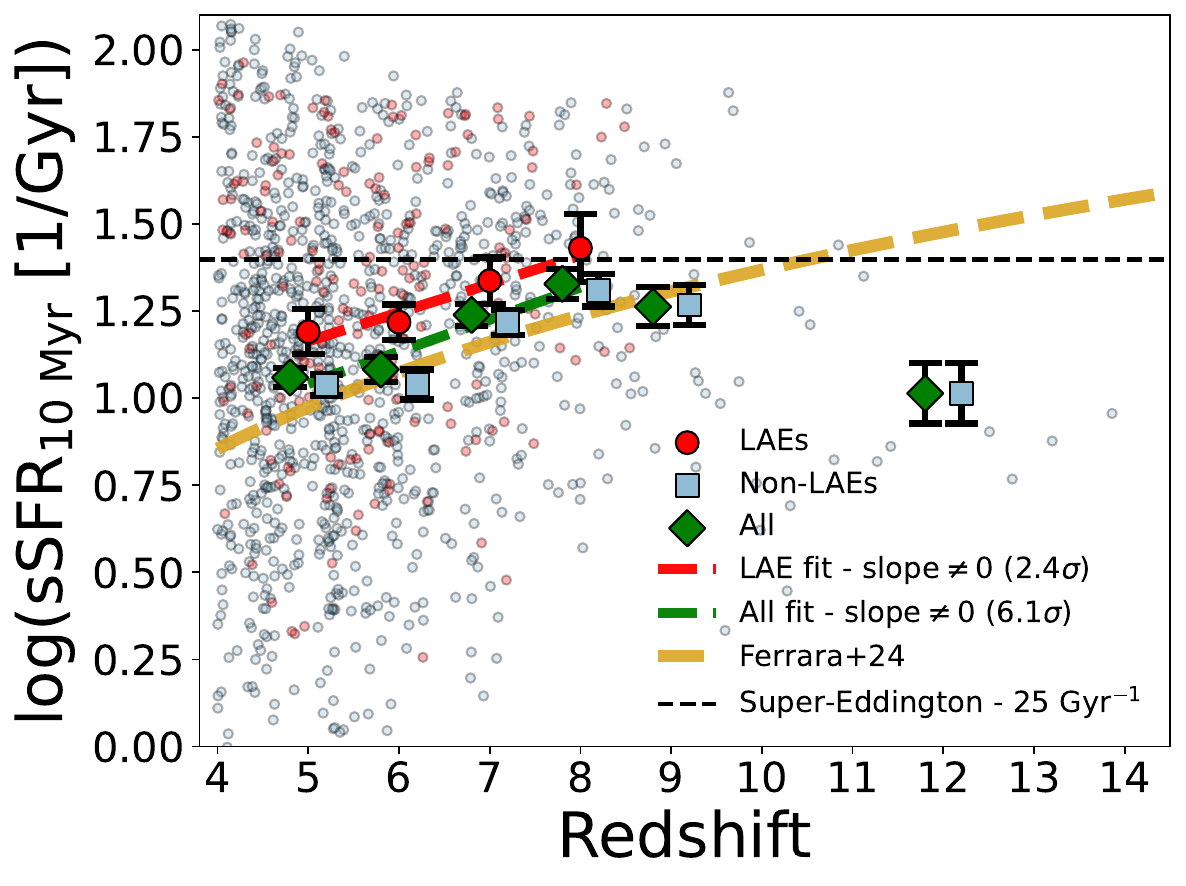}
    \end{subfigure}
    
    \vspace{0.05cm}

    \begin{subfigure}[b]{0.33\textwidth}
        \includegraphics[width=\linewidth]{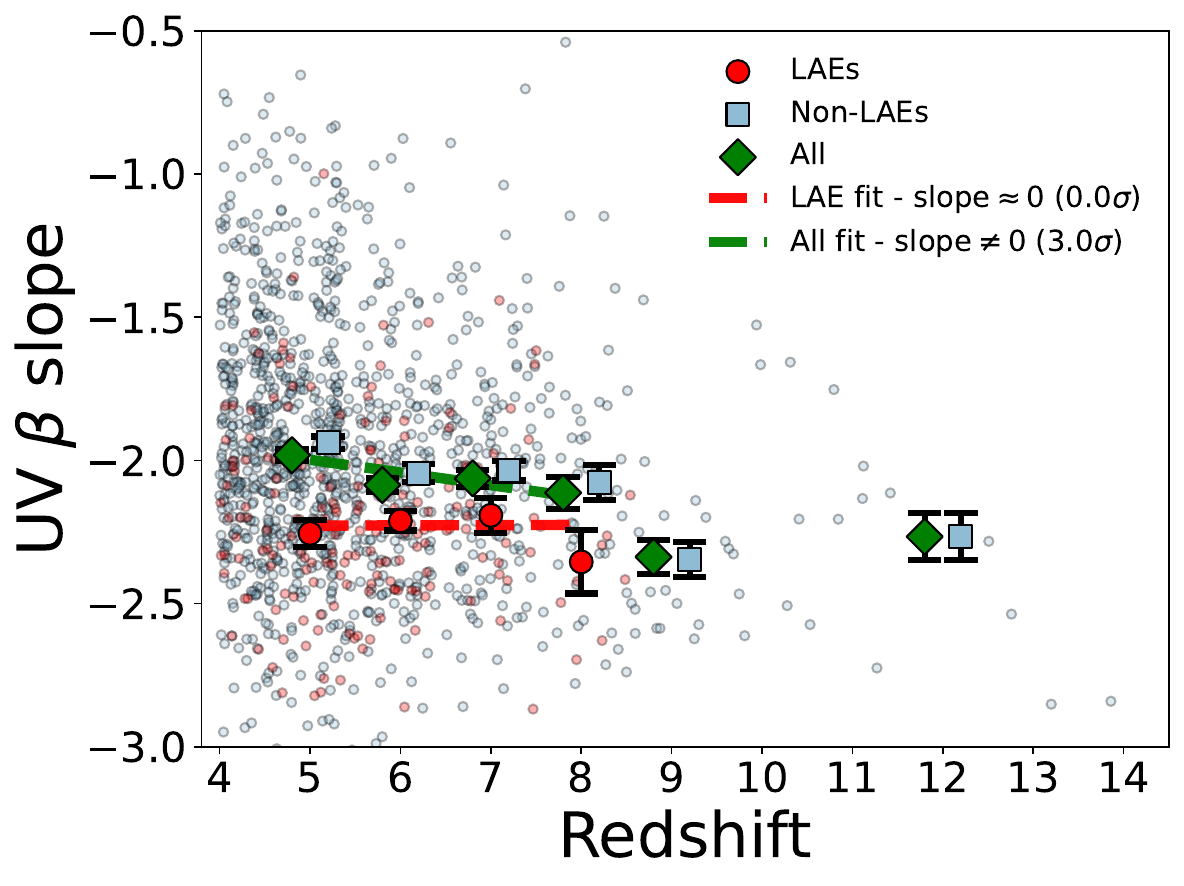}
    \end{subfigure}
    \begin{subfigure}[b]{0.33\textwidth}
        \includegraphics[width=\linewidth]{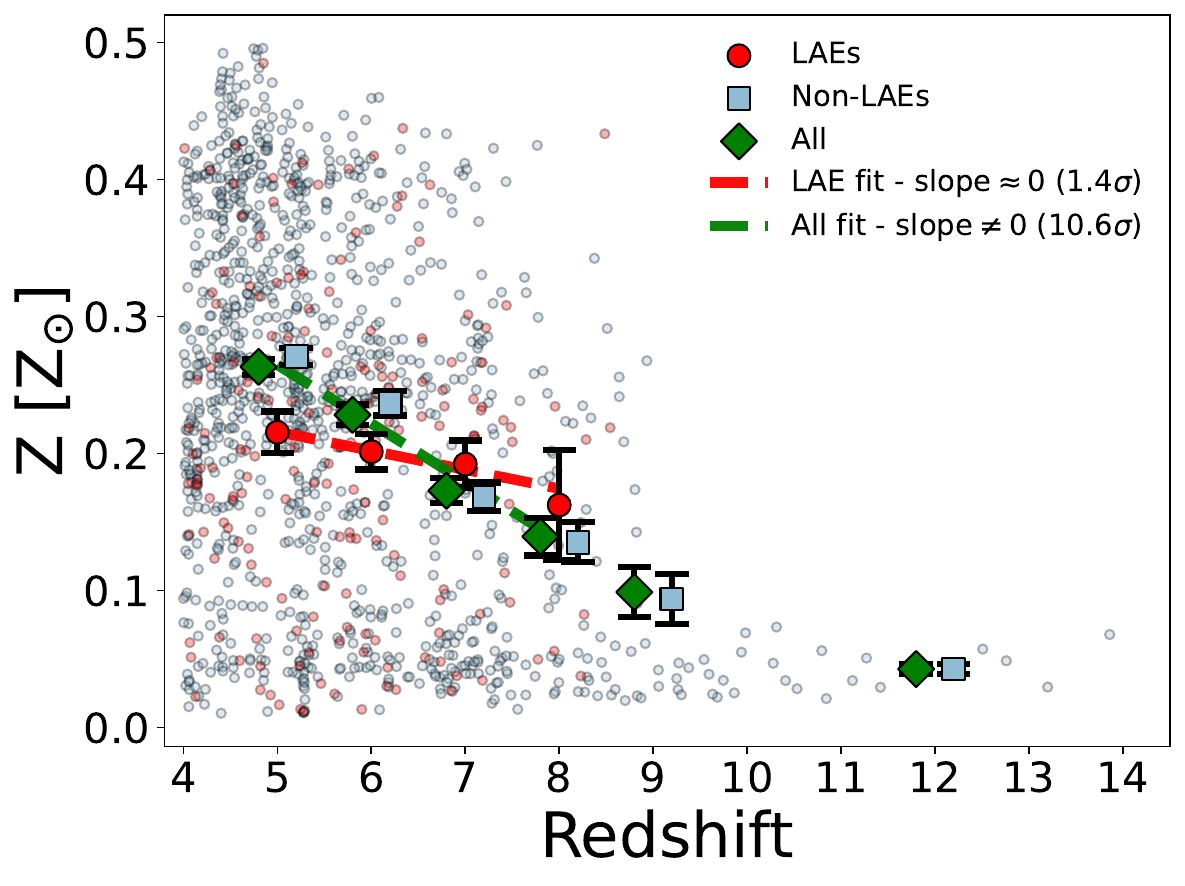}
    \end{subfigure}
    \begin{subfigure}[b]{0.33\textwidth}
        \includegraphics[width=\linewidth]{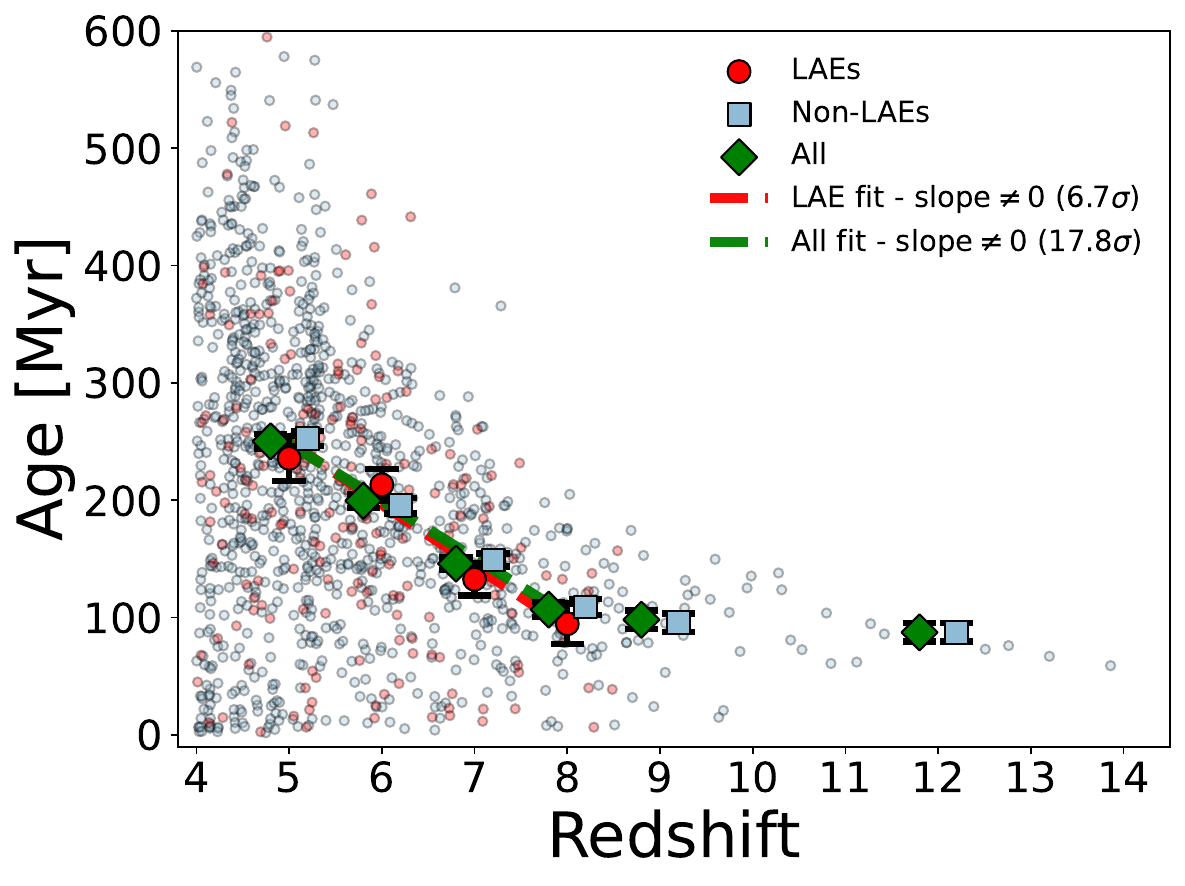}
    \end{subfigure}
    
    \caption{Redshift evolution of selected physical properties for galaxies in the intermediate UV-luminosity range, -20.25 < M$_{\mathrm{UV}}$ < -18.75. The panels show stellar mass, SFR$_{\textrm{30 Myr}}$, sSFR$_{\textrm{10 Myr}}$, UV slope $\beta$, metallicity, and mass-weighted age as a function of redshift. Red and blue circles in the foreground show individual LAEs and Non-LAEs, respectively. In each redshift bin, large red circles, cyan squares, and green diamonds mark the mean values for LAEs, Non-LAEs, and all SFGs, with corresponding uncertainties. Red and green dashed lines show the best-fitting linear relations to the binned mean values of LAEs and all SFGs, respectively. The legend reports the significance with which each fitted slope differs from zero, corresponding to no redshift evolution. Source-by-source Spearman correlation tests are reported in Table~\ref{tab:Spearman}. As discussed in the text, LAEs and non-LAEs occupy different regions of these diagrams. Overall, LAE properties show no evidence for significant redshift evolution, while the full SFG population evolves toward the locus occupied by LAEs at increasing redshift. The top-right panel also shows the redshift evolution and super-Eddington threshold discussed by \cite{Ferrara24a}.}
    \label{fig:Redshift_correlations} 
\end{figure*}

Assuming that the spatial extent of diffuse \lya\ halos does not evolve strongly over the redshift range considered, as suggested by extrapolations of VLT-MUSE results at z $\simeq$ 3--6 \citep[e.g.,][]{Guo2024}, differential slit losses should not dominate the observed trend in the \jwst\ measurements. Under this assumption, two main physical explanations remain. First, as proposed by \cite{Kageura2025}, UV-bright and intermediate-luminosity LAEs at high redshift may produce sufficient ionizing radiation, through the combination of their UV luminosities and Lyman-continuum escape fractions, to generate large ionized regions around themselves \citep[e.g.,][]{Jung2022, Witstok2024, Witstok2025Nature, Meyer2026, Whitler2026}. These bubbles would reduce the impact of resonant IGM absorption and facilitate the transmission of \lya\ photons even at very high redshift. Alternatively, the intrinsic production and escape of \lya\ may increase toward earlier epochs because the physical properties of SFGs progressively become more favorable to \lya\ emission. In the following section, we investigate the galaxy evolution scenario. The role of ionized bubbles will be explored in a forthcoming study.

\subsection{A galaxy evolution driven explanation}\label{sec:properties}
In this section, we investigate the physical conditions that distinguish LAEs from the remaining star-forming galaxy population, hereafter referred to as Non-LAEs, and examine whether these differences persist toward increasingly high redshift. To ensure a robust comparison between the two populations, in this section we define as LAEs all SFGs with a secure \lya\ detection at S/N > 3 and \lya\ EW$_0$ > 25~\AA. We define as Non-LAEs those SFGs with no significant \lya\ detection and a stringent EW$_{\mathrm{0,lim}}$ < 25~\AA. This latter requirement minimizes contamination by objects for which the non-detection may simply result from insufficiently deep observations. 
Within the intermediate M$_{\mathrm{UV}}$ range, our sample comprises 195 LAEs and 1068 Non-LAEs. We also consider a more inclusive selection, defined by the 60\% relative completeness limit of the sample, M$_{\mathrm{UV}}$ < -18, which includes 358 LAEs and 1815 Non-LAEs. As previously discussed, the observed \lya\ EW$_0$ correlates with M$_{\mathrm{UV}}$, while several ISM and stellar-population properties also depend on UV luminosity. Restricting the comparison to a well-defined M$_{\mathrm{UV}}$ range is therefore essential for obtaining a meaningful comparison between the two populations.

We first perform a two-sample Kolmogorov-Smirnov test on the properties of LAEs and Non-LAEs derived from the SED-fitting analysis described in Sect.~\ref{sec:SED_fitting}, namely stellar mass, SFR, E(B-V), Z, age, together with the spectroscopically measured UV slope $\beta$ described in Sect.~\ref{sec:Muv}. We further define the specific star formation rate (sSFR) as the ratio between the SFR in the last 10~Myr and the stellar mass, and the burstiness parameter as the ratio between the SFR averaged over the most recent 10 and 30~Myr. For both adopted M$_{\mathrm{UV}}$ selections, we obtain p-values $<<$ 5$\times$10$^{-2}$, for the stellar mass, SFR, sSFR, burstiness, stellar reddening, metallicity, and UV slope $\beta$ distributions. We therefore reject the null hypothesis that the LAE and Non-LAE distributions of these properties are drawn from the same parent population. In the intermediate M$_{\mathrm{UV}}$ selection, the median stellar mass, SFR, stellar reddening, and metallicity of LAEs are lower than those of non-LAEs by factors of 2.4, 1.8, 1.8, and 1.2, respectively. LAEs also have bluer UV slopes, with a median offset of $\Delta\beta$ = 0.2, while their median sSFR and burstiness are higher by factors of 1.5 relative to Non-LAEs. Instead, for the mass-weighted age, we do not find statistically significant evidence of a difference between the two populations (p-value $\sim$ 0.1). These results are consistent with previous studies \citep[e.g.,][and references therein]{Gawiser2006, Ouchi_2020, Napolitano2023, Laferte-Urrutia2026}, which have shown that LAEs typically have lower mass and less dust than Non-LAEs and are more likely to have experienced a recent burst of star formation than the average SFG population at similar stellar masses. These properties can enhance both the intrinsic production and the escape of \lya\ photons. In particular, recent star formation episodes increase the production of ionizing photons and hence the recombination-powered \lya\ emission, while a lower dust content in the ISM facilitates the escape of resonantly scattered \lya\ radiation \citep[e.g.,][]{Saxena2024, Napolitano2024, Almada-Monter2026}.

Fig.~\ref{fig:Redshift_correlations} presents the properties described above for individual LAEs and Non-LAEs as a function of redshift. Using the same redshift bins adopted for the X$_{\mathrm{Ly\alpha}}$ measurement, we also show the mean values and corresponding uncertainties for the LAE, Non-LAE, and full SFG populations. In each redshift interval, we report a binned measurement only when the corresponding population includes more than three objects. As a result, the mean values for the LAE population are shown only up to z $\leq$ 8.5. We find that the attenuation-free model \citep[AFM;][]{Ferrara23, Ferrara24a, Ferrara26} provides a good description of the observed trend up to z = 8--9 \citep[see also][]{Santini2026}.

We next investigate the redshift evolution of the properties of LAEs and compare it with that of the SFG population. For each property, we first performed a Spearman rank-correlation test with redshift using individual sources. We consider a correlation to be statistically significant when the p-value is less than 0.05. For the LAE population, we find no statistically significant redshift evolution in any of the considered properties, with the exception of mass-weighted age, which decreases with increasing redshift, as expected from the progressively younger cosmic age at earlier epochs. This result is recovered for both the intermediate and completeness-limited M$_{\mathrm{UV}}$ selections. The absence of significant evolution suggests that observable LAEs occupy an extreme and stable region of galaxy-property space over the redshift range probed by our sample.\\
We then repeated the same analysis for the full SFG population. In this case, we find statistically significant anticorrelations between redshift and stellar mass, UV slope $\beta$, stellar reddening, SFR, metallicity, and mass-weighted age. Instead, sSFR and burstiness are positively correlated with redshift. These trends are recovered in both the completeness-limited and intermediate M$_{\mathrm{UV}}$ selections. Spearman rank-correlation results are summarized in Table~\ref{tab:Spearman}. 
We emphasize that the Spearman analysis is performed on the individual sources rather than on the binned mean values. This choice preserves the intrinsic scatter of the galaxy population and provides a more representative assessment of the overall redshift trends in the current dataset. \\ 
As an additional consistency check, we fitted the binned mean values of each physical property with a linear relation, using inverse-variance weights and restricting the fit to the redshift range where binned measurements are defined for both LAEs and SFGs (i.e., z $\leq$ 8.5). The linear relation is not intended as a physical model for the redshift evolution, but as a diagnostic of whether the binned trends retain a slope significantly different from zero. The best fitting linear relations for the LAE and full SFG population are shown in Fig.~\ref{fig:Redshift_correlations}, where we indicate whether each best-fit slope is consistent with no redshift evolution, corresponding to a null slope. For the full SFG population, all properties showing a significant Spearman correlation or anticorrelation also have fitted slopes that differ from zero at more than 3$\sigma$, independently confirming the trends recovered from the source-by-source Spearman analysis. Conversely, for the LAE population, for which no significant redshift evolution is recovered from the Spearman test, except for the mass-weighted age, all fitted slopes are consistent with zero within 2$\sigma$. The only exceptions are the already discussed evolution of the mass weighted age, whose slope differs from zero at 6.7$\sigma$, and a tentative evolution of the sSFR, whose slope differs from zero at 2.4$\sigma$.

Taken together, our results indicate that, at progressively higher redshifts, the parent SFG population shifts toward the region of stellar and ISM properties occupied by LAEs at all epochs. Therefore, galaxy evolution is expected to create increasingly favorable conditions that enhance the intrinsic production and escape of \lya\ photons, shifting the intrinsic \lya\ EW$_0$ probability distribution toward larger values. In the absence of attenuation by the IGM, this evolution would naturally produce an increase in \lya\ visibility with redshift. The observed \lya\ EW$_0$ distribution, and consequently X$_{\mathrm{Ly\alpha}}$, is also modulated by the evolving IGM transmission, which increasingly suppresses the visibility of \lya\ emission during the EoR (Sect.~\ref{sec:XLya}). Many previous studies, including our own \citep[e.g.,][and references therein]{Pentericci_2018b, Napolitano2026}, have neglected the evolution of the underlying galaxy population when inferring the IGM neutral-hydrogen fraction, effectively assuming that the intrinsic production and escape of \lya\ remain unchanged from z $\simeq$ 6 and above. Our findings call this assumption into question and highlight the need to account explicitly for galaxy evolution in future analyses. In the following section, we discuss how the observed \lya\ visibility can be used to constrain the progress of cosmic reionization, while accounting for the underlying evolution of the galaxy population.

\begin{table}
\caption{Spearman rank-correlation analysis between redshift and the physical properties of LAEs and the full SFG population.}
\label{tab:Spearman}
\centering
\begin{tabular}{ccccc}
\hline \noalign{\smallskip}
& \multicolumn{4}{c}{-20.25 < M$_{\mathrm{UV}}$ < -18.75} \\
\noalign{\smallskip}
& \multicolumn{2}{c}{LAEs} & \multicolumn{2}{c}{All SFGs} \\
\noalign{\smallskip}
Observable & $\rho_{\mathrm{S}}$ & p-value & $\rho_{\mathrm{S}}$ & p-value \\
\hline
\noalign{\smallskip}
Stellar mass & -- & 0.2 & -0.2 & $\ll$ 0.05 \\ [3pt]
UV slope $\beta$ & -- & 0.4 & -0.2 & $\ll$ 0.05 \\ [3pt]
E(B-V) & -- & 0.6 & -0.2 & $\ll$ 0.05 \\ [3pt]
SFR$_{\mathrm{10Myr}}$ & -- & 0.4 & -0.2 & $\ll$ 0.05 \\ [3pt]
SFR$_{\mathrm{10Myr}}$/SFR$_{\mathrm{30Myr}}$ & -- & 0.9 & 0.06 & 0.04 \\ [3pt]
sSFR & -- & 0.7 & 0.06 & 0.02 \\ [3pt]
Z & -- & 0.6 & -0.3 & $\ll$ 0.05 \\ [3pt]
Age & -0.3 & $\ll$ 0.05 & -0.3 & $\ll$ 0.05 \\ [3pt]
\hline
\end{tabular}
\tablefoot{Correlations are considered statistically significant when p < 0.05. A dash indicates that the correlation coefficient is not reported for a non-significant correlation.}
\end{table}

\subsection{A model for the evolution of the \lya\ fraction}\label{sec:AFM_pred}

To assess whether the observed evolution of the \lya\ fraction can be understood within a physically motivated framework, we constructed a simple semi-empirical model based on the AFM \citep[][]{Ferrara23, Ferrara24a, Ferrara26}, linking the intrinsic properties of high-redshift galaxies to the transmission of \lya\ photons through the evolving IGM. The model is intentionally minimal and is designed to capture the dominant physical processes while introducing only a small number of free parameters. A full description of the model will be given elsewhere (Ferrara et al., in prep.). Below, we briefly summarize its main features. 

At each redshift, the parent galaxy population is described by the observed UV luminosity function, which provides the number density of galaxies within the UV-magnitude interval considered in the analysis. UV luminosities are converted into star formation rates, assuming a standard UV$-$SFR calibration, which we take from the AFM:
\begin{equation}
M_{\rm UV} = 5.89 - 2.5 \log_{10}\!\left(\kappa_{1500}\,{\rm SFR}\right),
\end{equation}
with
\begin{equation}
\kappa_{1500} = 0.587\times 10^{10}\,
\frac{L_\odot}{M_\odot\,{\rm yr}^{-1}}.
\end{equation}
We also assume a redshift-dependent photon production efficiency, adopting the empirical relation from \cite{Llerena2025}, $\log_{10} \xi_{\rm ion} = 24.82+0.06 \ z$. We do not include an explicit dependence of $\xi_{\rm ion}$ on M$_{\rm UV}$, since this dependence is relatively weak over the intermediate UV-luminosity range considered, i.e., -20.25 < M$_{\mathrm{UV}}$ < -18.75, and remains within the range of values already explored by our adopted model (see Fig.~5 of \citealt{Llerena2025} and Fig.~8 of \citealt{Begley2025}). In Sect.~\ref{sec:AFM_const_xion}, we further explore how our results change when adopting, instead, the canonical constant value of $\log_{10} \xi_{\rm ion}$ = 25.2.\\ 
The ionizing photon production rate is
\begin{equation}
\dot N_{\rm ion} = \xi_{\rm ion}\, L_{\nu,{\rm UV}}.
\end{equation}
The production rate of Ly$\alpha$ photons is then computed assuming Case-B recombination.

The escape of Ly$\alpha$ photons from the ISM of galaxies is modeled using radiative transfer calculations from \cite{Orsi2012}, in which the escape fraction depends on the neutral hydrogen column density and the outflow velocity of the gas. Rather than adopting a single outflow velocity, we assume that galaxies follow the AFM prediction for the sSFR distribution. The outflow velocity is linked to the Eddington ratio, $\ell={\rm sSFR}/{\rm sSFR}_*$, where ${\rm sSFR}_*\approx 25\, {\rm Gyr}^{-1}$ (dashed horizontal line in the lower left panel of Fig.~\ref{fig:Redshift_correlations}), through a simple prescription in which quiescent galaxies exhibit negligible outflows, while rapidly star-forming systems develop increasingly fast winds (see Sec. 4 and eq. 15 of \citealt{Ferrara24a}, see also \citealt{Fiore2023}). The outflow velocity (v$_{\mathrm{0}}$) of a galaxy at the Eddington ratio threshold ($\ell \equiv 1$) is left as a free parameter of the model. The intrinsic probability that a galaxy is observed as a Ly$\alpha$ emitter is therefore obtained by integrating over the full sSFR distribution.

The intrinsic Ly$\alpha$ equivalent width is subsequently attenuated by resonant Gunn-Peterson scattering in the neutral IGM. The effective optical depth is assumed to scale with the volume-averaged neutral hydrogen fraction x$_{\mathrm{HI}}$, the expansion velocity of the Ly$\alpha$ line, and redshift. Several reionization histories are explored, including published trends from \citet[][]{Ishigaki2018, Finkelstein2019, Naidu2020, Yung2020b, Qin2025, Kageura2026}. This allows us to isolate the sensitivity of the predicted Ly$\alpha$ fraction to the assumed evolution of the neutral IGM.

The predicted Ly$\alpha$ fraction is finally computed as
\[
X_{\rm Ly\alpha}(z)=
\frac{N_{\rm LAE}(z)}
     {N_{\rm gal}(z)},
\]
where the numerator is obtained by integrating the UV luminosity function weighted by the probability that each galaxy satisfies the adopted equivalent-width threshold of 25 \AA, while the denominator is simply the total number density of UV-selected galaxies over the same magnitude interval of interest. In the next section, we adopt -20.25 < M$_{\rm UV}$ < -18.75 to directly compare with the observed \lya\ fraction results presented in Sect.~\ref{sec:XLya} in a fully self-consistent manner.

\subsection{\lya\ constraints on reionization}\label{sec:reionization}
\begin{figure*}[ht!]
    \centering

    \begin{subfigure}[b]{0.49\textwidth}
        \includegraphics[width=\linewidth]{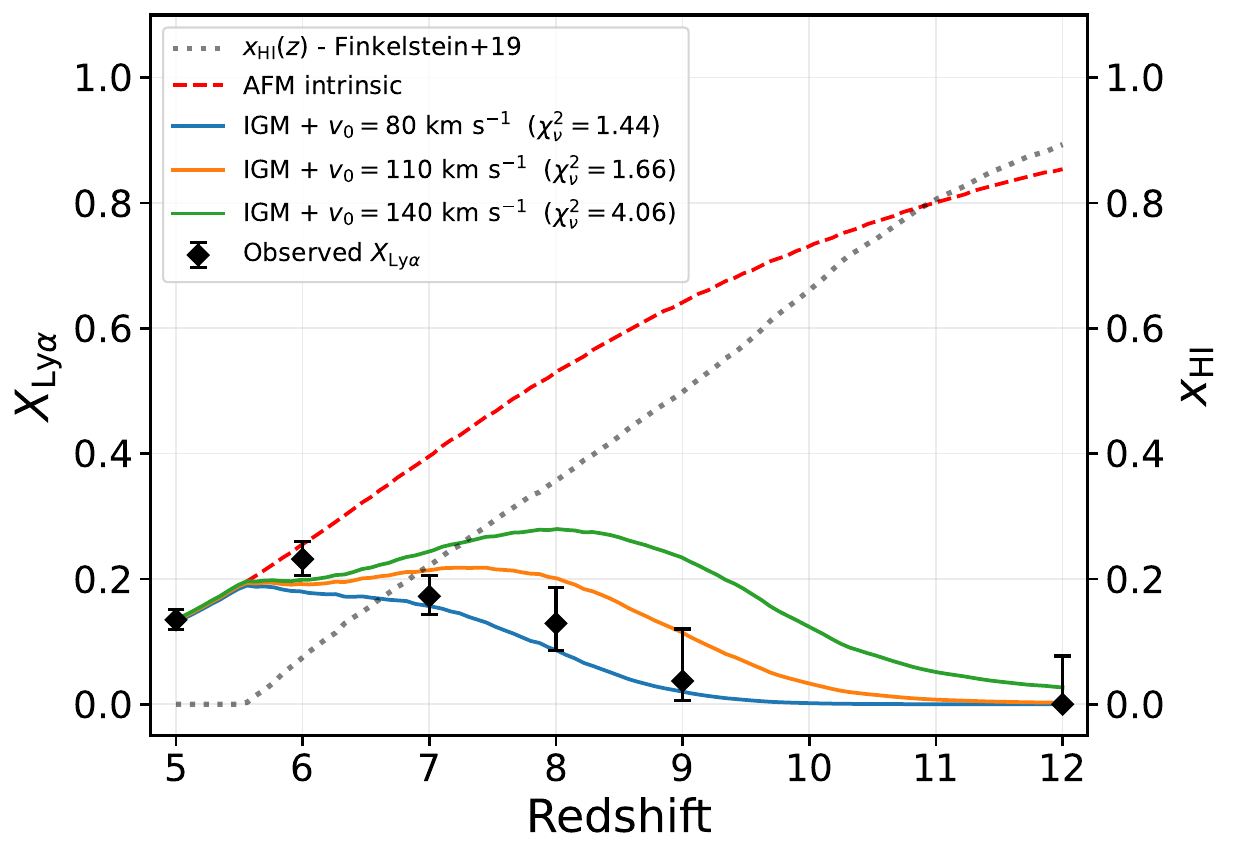}
    \end{subfigure}
    \begin{subfigure}[b]{0.49\textwidth}
        \includegraphics[width=\linewidth]{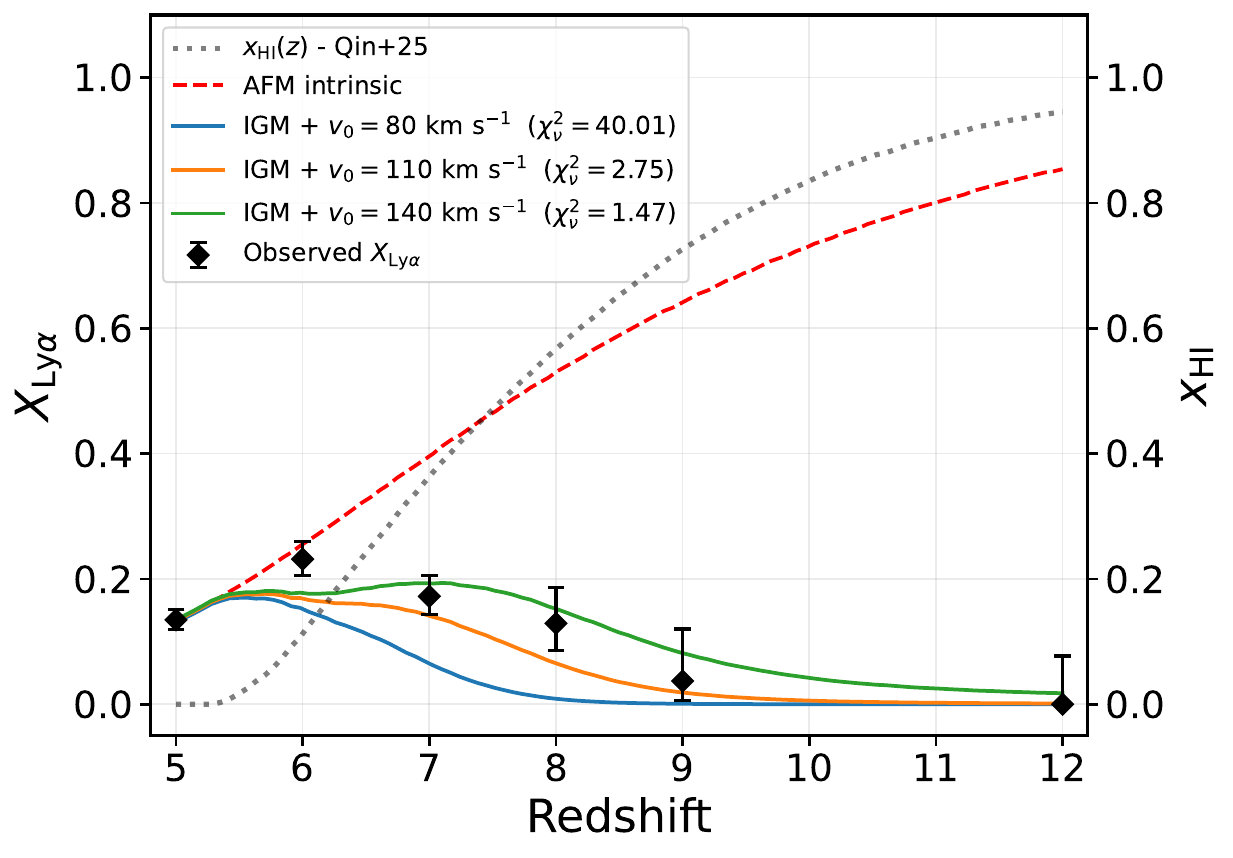}
    \end{subfigure}
    
    \vspace{0.01cm}

    \begin{subfigure}[b]{0.49\textwidth}
        \includegraphics[width=\linewidth]{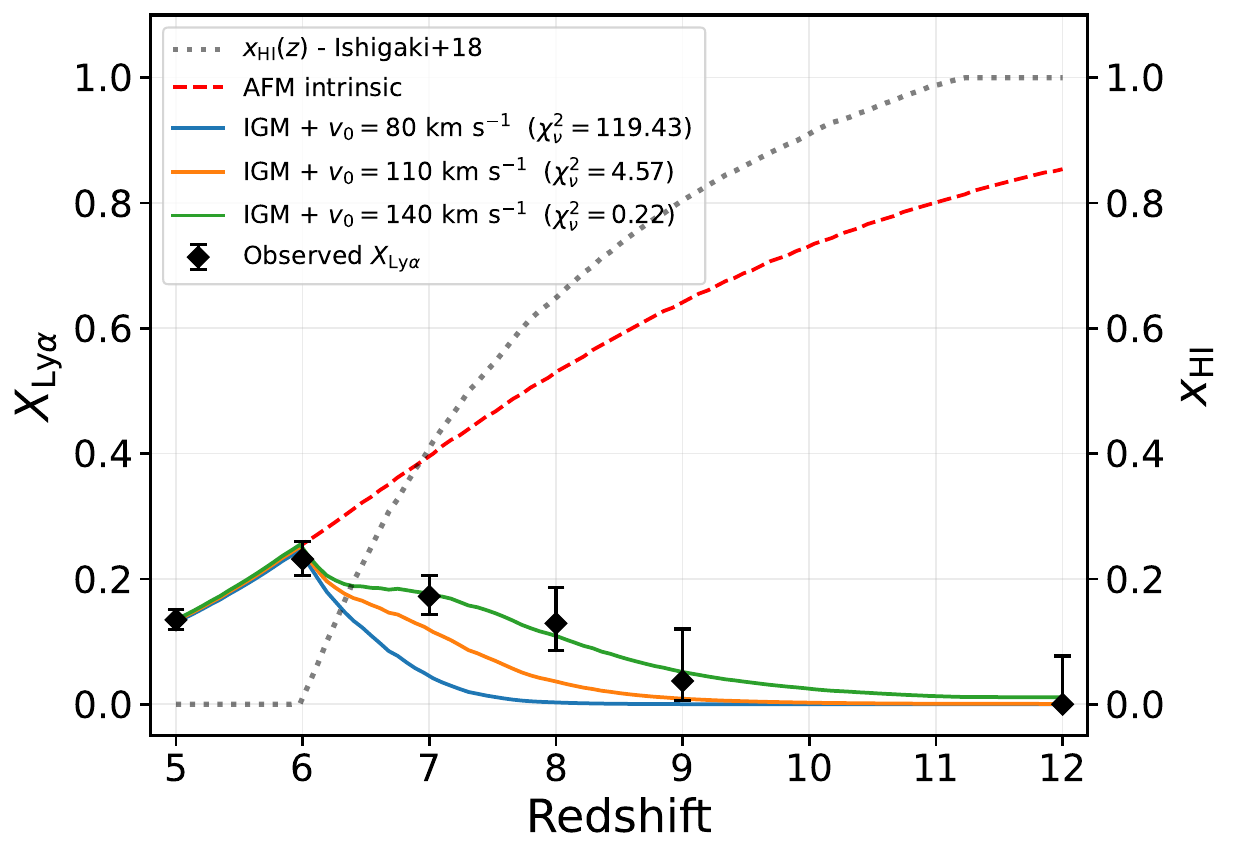}
    \end{subfigure}
    \begin{subfigure}[b]{0.49\textwidth}
        \includegraphics[width=\linewidth]{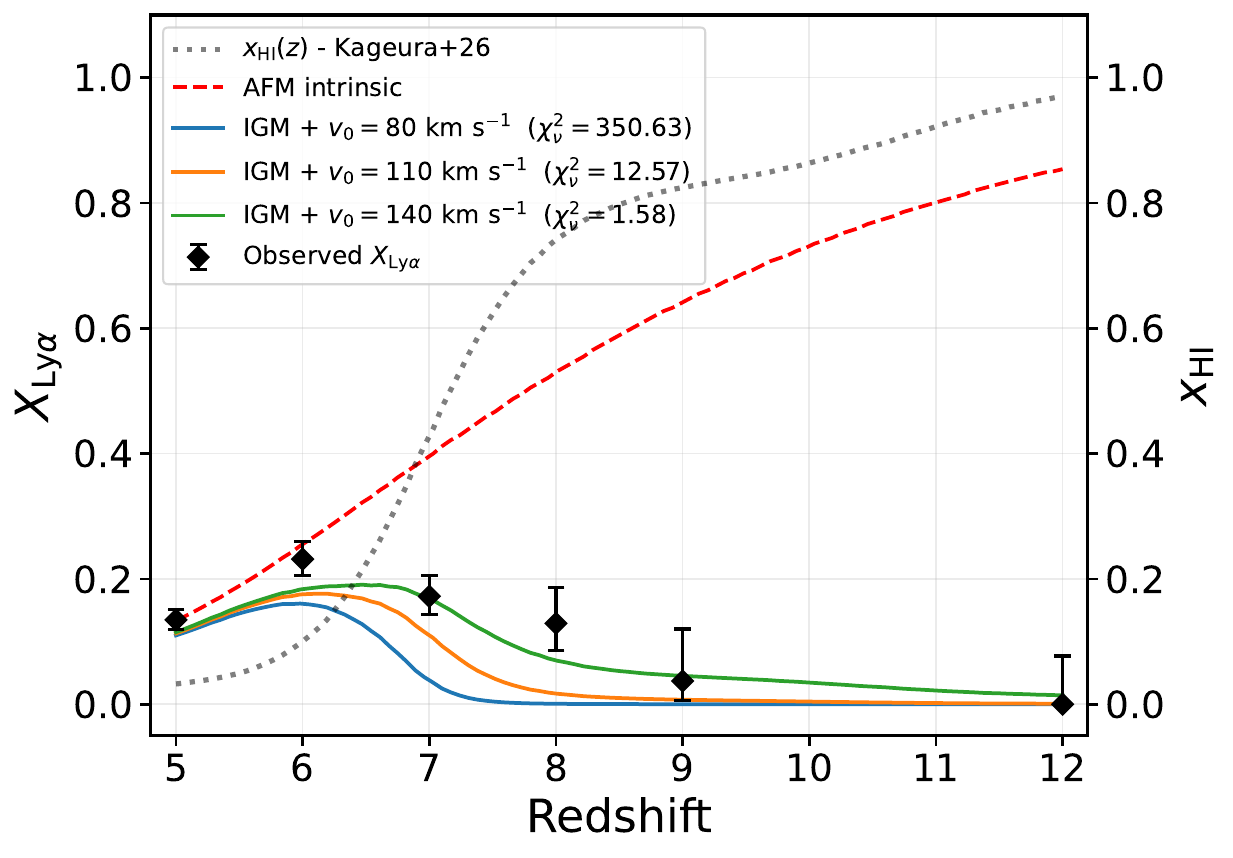}
    \end{subfigure}

    \vspace{0.01cm}

    \begin{subfigure}[b]{0.49\textwidth}
        \includegraphics[width=\linewidth]{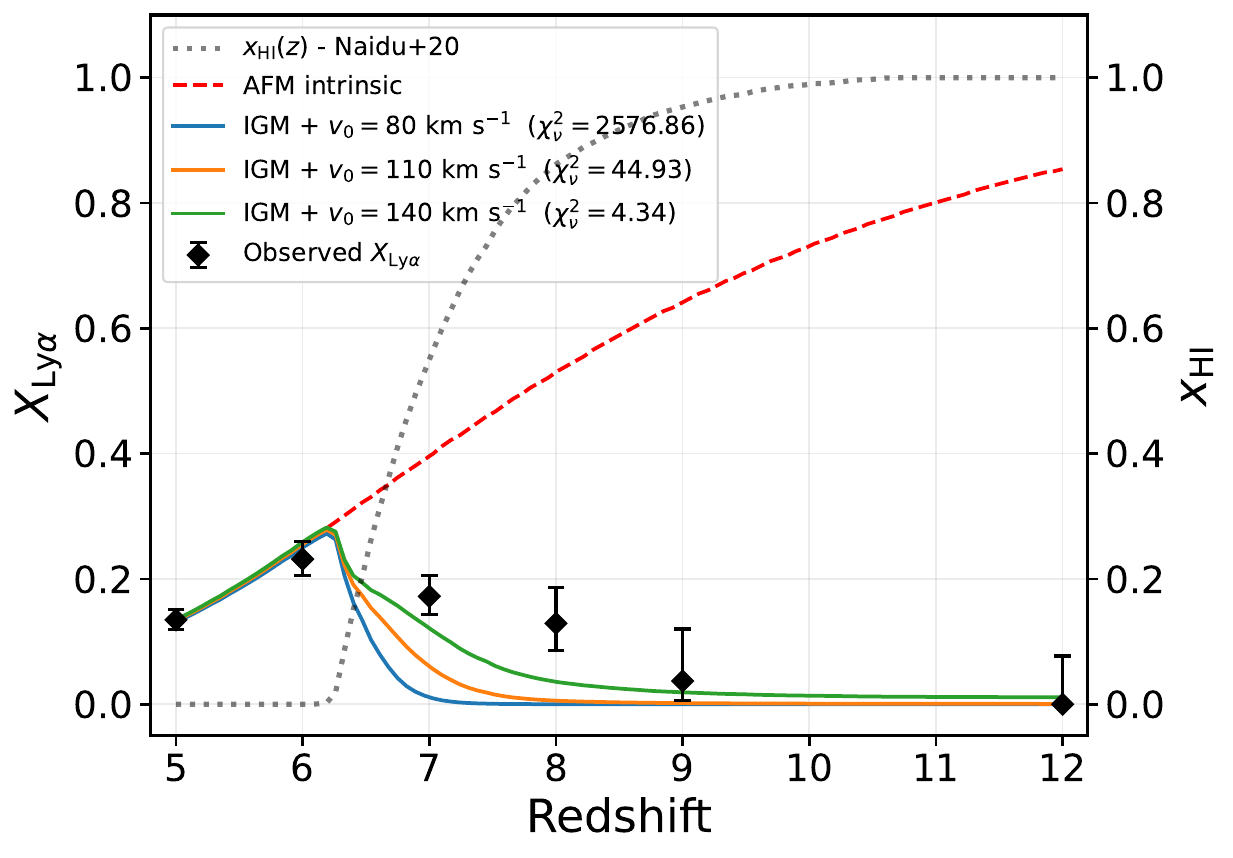}
    \end{subfigure}
    \begin{subfigure}[b]{0.49\textwidth}
        \includegraphics[width=\linewidth]{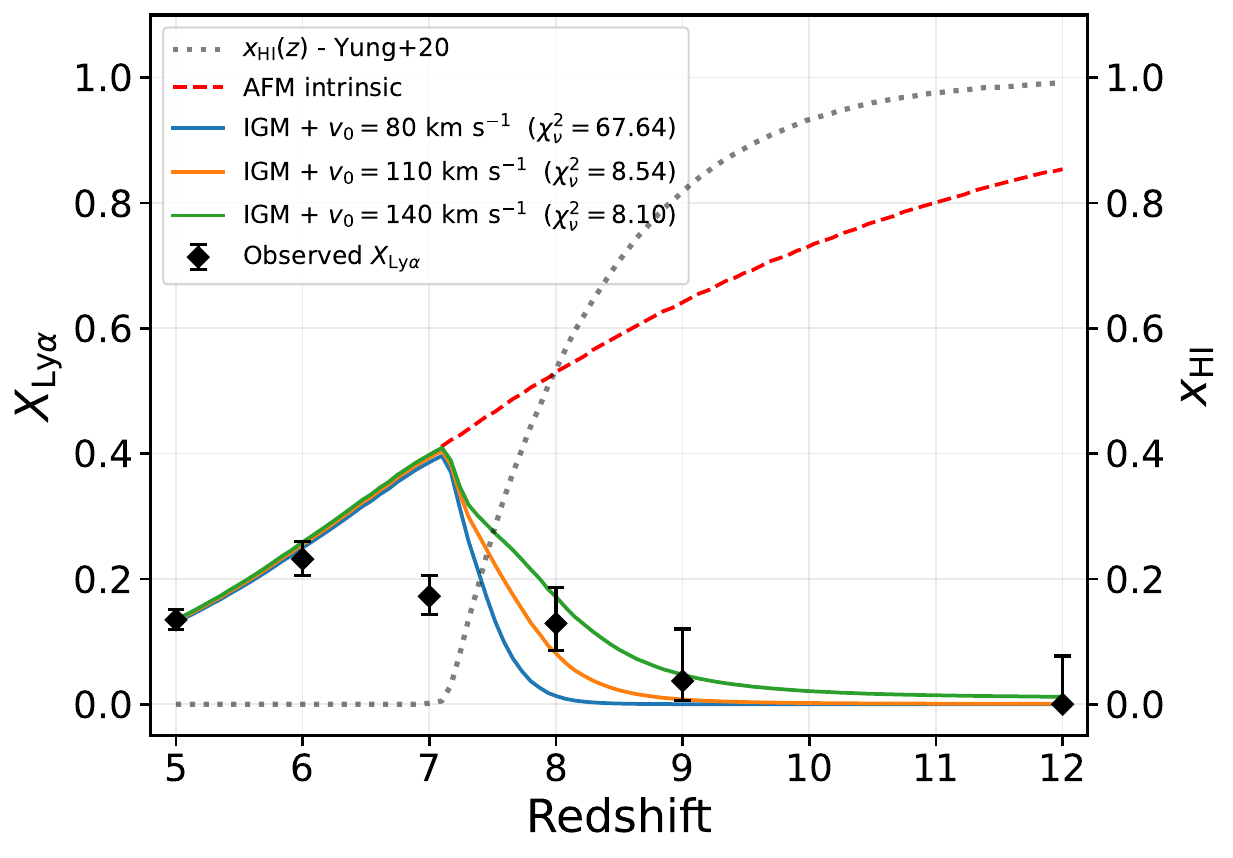}
    \end{subfigure}
    
    \caption{Model prediction of X$_{\mathrm{Ly\alpha}}$ accounting for both IGM and galaxy evolution. The left vertical axis refers to X$_{\mathrm{Ly\alpha}}$, while the right vertical axis indicates x$_{\mathrm{HI}}$. The six panels differ in the adopted reionization history, which determines the evolution of x$_{\mathrm{HI}}$, shown by the gray dotted curves. The top row shows relatively shallow reionization histories \citep{Finkelstein2019, Qin2025}; the middle row presents models characterized by slow initial and final evolution separated by a rapid intermediate phase \citep{Ishigaki2018, Kageura2026}; and the bottom row shows late-rapid and early-rapid reionization scenarios \citep{Naidu2020, Yung2020b}. The red dashed curves represent the same intrinsic evolution of X$_{\mathrm{Ly\alpha}}$ driven by galaxy evolution alone, as predicted by the AFM assuming an outflow velocity at the Eddington ratio threshold of v$_{\mathrm{0}}$ = 110 km/s. For each panel, the three solid curves show the combined effects of galaxy evolution and IGM transmission, following Ferrara et al. (in prep.), and are color-coded according to the assumed v$_{\mathrm{0}}$, treated as a free parameter of the model \citep{Ferrara2024}. Black diamonds show the observed \lya\ fractions obtained in this work.}
    \label{fig:EoR_hystories} 
\end{figure*}
We compare the observed X$_{\mathrm{Ly\alpha}}$ measurements obtained in Sect.~\ref{sec:XLya} with the semi-empirical model of galaxy evolution based on the AFM. Our aim is to highlight the combined effects of the evolving galaxy population and IGM attenuation on \lya\ visibility (Fig.~\ref{fig:EoR_hystories}).

The intrinsic AFM prediction is anchored to the observed X$_{\mathrm{Ly\alpha}}$ measurement at z = 5. We find that the subsequent evolution driven by galaxy evolution alone, for physically plausible values of the characteristic outflow velocity, neutral-hydrogen column density, and characteristic sSFR that regulate the scattering and escape of \lya\ photons through the ISM \citep[e.g.,][]{Dijkstra2014, Mason2018a}, also reproduces the observed X$_{\mathrm{Ly\alpha}}$ value at z = 6. This is consistent with the expectation that IGM attenuation remains weak up to z $\sim$ 6, where quasar-based constraints indicate very low neutral fractions, x$_{\rm HI}$ $\sim$ 10$^{-3}$--10$^{-2}$ \citep[e.g., see Fig.~2 of][for a recent review]{Ellis2025}. In this regime, X$_{\mathrm{Ly\alpha}}$ is therefore expected to be only weakly sensitive to such small values of x$_{\rm HI}$.
At higher redshifts, during the EoR, the additional resonant scattering of \lya\ photons by neutral hydrogen in the IGM must be taken into account. The six panels in Fig.~\ref{fig:EoR_hystories} explore six reionization histories drawn from recent studies. The upper panels correspond to scenarios in which reionization begins early and proceeds gradually \citep{Finkelstein2019, Qin2025}, the middle row considers histories characterized by a more pronounced rapid intermediate transition preceded by slower evolution \citep{Ishigaki2018, Kageura2026}, while the bottom row shows more rapid reionization histories, starting either at later or earlier times \citep{Naidu2020, Yung2020b}. In each case, the same intrinsic evolution of \lya\ visibility is combined with the corresponding evolution of the IGM neutral fraction.

We quantified the agreement between the predicted and observed LAE fractions using a Pearson $X^2$ statistic applied directly to the binomial counts in each redshift bin. Results are reported in each panel of Fig.~\ref{fig:EoR_hystories}. The comparison reveals a clear preference for models in which the neutral fraction increases gradually over the redshift interval probed by our measurements. The best agreement is obtained for the \cite{Ishigaki2018} history combined with v$_{\mathrm{0}}$ = 140~km~s$^{-1}$. The \cite{Finkelstein2019} history also provides a good description of the observations for relatively small characteristic outflow velocities, v$_{\mathrm{0}}$ = 80--110~km~s$^{-1}$. Similarly, the \cite{Qin2025} and \cite{Kageura2026} histories remain consistent with the observations when combined with a larger characteristic velocity of v$_{\mathrm{0}}$ = 140~km~s$^{-1}$. Thus, while the detailed preference among the more gradual reionization histories remains degenerate with the assumed \lya\ velocity offset, several of these scenarios can reproduce the observed evolution in the physically motivated framework described in Sect.~\ref{sec:AFM_pred}.

By contrast, the more rapid reionization histories considered are strongly disfavored over the full range of adopted outflow velocities, regardless of whether they started earlier \citep{Yung2020b} or at later times \citep{Naidu2020}. The tension is driven primarily by the relatively gradual decline of the observed X$_{\mathrm{Ly\alpha}}$ between z = 6 and z = 8: a rapid increase in x$_{\mathrm{HI}}$ over this interval suppresses the predicted \lya\ visibility more abruptly than indicated by the observations. \\

We stress that our comparison between different reionization histories and current observed X$_{\mathrm{Ly\alpha}}$ constrains is degenerate with the assumed normalization of the outflow velocities, v$_{\mathrm{0}}$. 
This degeneracy is directly apparent in Fig.~\ref{fig:EoR_hystories}: histories that provide poor fits for small v$_{\mathrm{0}}$ can become fully consistent with the observations when larger velocity offsets are adopted. Breaking this degeneracy will require systematic measurements of \lya\ velocity offsets over a broad redshift range. Such measurements will rely on large samples observed at higher spectral resolution with the \jwst\ NIRSpec G140M and G140H dispersers rather than the \NIRSpec\ PRISM configuration.

\section{Summary} \label{sec:Conclusion}
In this work, we investigated the redshift evolution of \lya\ visibility and the ISM properties of galaxies across the five CANDELS fields using publicly available \jwst/\NIRSpec\ PRISM spectroscopy. Our analysis spans 4 $\leq $ z < 14.2 and includes 3446 unique sources. We classified 3361 sources as SFGs and identified the remaining 85 sources as BLAGNs. Within the SFG population, 539 galaxies exhibit secure \lya\ emission detected at S/N > 3. The parent sample is complete down to a 60\% relative completeness limit of M$_{\mathrm{UV}}$ = -18.

For the subsequent analysis, we defined LAEs as SFGs with a secure (S/N > 3) \lya\ detection and rest-frame EW$_0$ > 25~\AA. Non-LAEs were instead selected among SFGs without significant \lya\ emission and whose spectra were sufficiently sensitive to reach EW$_{0,\mathrm{lim}}$ < 25~\AA. Based on these definitions, our main results can be summarized as follows:

\begin{itemize}
\item We measured the observed \lya\ emitter fraction, X$_{\mathrm{Ly\alpha}}$, averaged over all five CANDELS fields across the redshift interval 4 $\leq $ z < 14.5. The analysis was performed in two absolute UV magnitude ranges of -20.25 < M$_{\mathrm{UV}}$ < -18.75 and -21.75 < M$_{\mathrm{UV}}$ < -20.25, which are well above the completeness limit of M$_{\mathrm{UV}}$ = -18. In the fainter UV range, X$_{\mathrm{Ly\alpha}}$ increases from z = 5 to 6 and then declines toward higher redshifts. The decline at z > 6 is supported by a significant monotonic trend.
In the brighter UV-luminosity range, although X$_{\mathrm{Ly\alpha}}$ shows a tentative decline at z > 7, the small number of detected LAEs prevents us from establishing a statistically significant redshift evolution.
\item Using two-sample Kolmogorov-Smirnov tests, we confirmed that LAEs and Non-LAEs occupy distinct regions of galaxy-property space. 
In the intermediate M$_{\mathrm{UV}}$ selection, the median stellar mass, stellar reddening, SFR, and metallicity of LAEs are lower than those of Non-LAEs.
LAEs also have bluer UV slopes $\beta$,
while their median sSFR and burstiness are higher.
These differences suggest that observable \lya\ emission is preferentially observed in the lower-mass, less attenuated, more metal-poor, and more bursty galaxy population.
\item We investigated the redshift evolution of the physical properties of LAEs using source-by-source Spearman rank-correlation tests, complemented by linear fits to the binned mean trends. None of the considered properties show a statistically significant correlation with redshift, with the exception of mass-weighted age, which is anticorrelated with increasing redshift, 
as expected from the progressively younger age of the Universe at earlier epochs. The best-fit slope analysis also supports this result: all LAE slopes are consistent with no redshift evolution,
except for mass-weighted age, whose slope significantly differs from zero, 
and a tentative evolution of sSFR. 
These results suggest that observable LAEs occupy an extreme and comparatively stable region of galaxy-property space over the redshift range probed by our sample.
\item We performed the same source-by-source Spearman correlation analysis for the full SFG population and found statistically significant redshift evolution in all the considered properties.
Toward higher redshifts, SFGs are characterized by lower stellar masses, bluer UV slopes, lower stellar reddening, lower SFR$_{\mathrm{10\ Myr}}$, and lower metallicities, together with higher sSFRs and burstiness. These trends are independently confirmed by the linear fits to the binned mean values, whose slopes significantly differ from zero
for all properties showing significant Spearman correlations. These results indicate that, at progressively higher redshifts, the parent SFG population increasingly overlaps with the region of stellar and ISM-property space occupied by LAEs at all epochs. Galaxy evolution should therefore enhance both the intrinsic production and escape of \lya\ photons, shifting the intrinsic \lya\ EW$_0$ distribution and consequently increasing the \lya\ emitter fraction expected in the absence of IGM attenuation at higher redshifts.
\item At variance with assumptions commonly adopted in previous studies, we concluded that both galaxy evolution and IGM attenuation must be considered when using \lya\ visibility to infer the cosmic neutral hydrogen fraction, x$_{\mathrm{HI}}$. In this context, we explored a physically motivated framework (\cite{Ferrara24a}, Ferrara et al. in prep.) that links the intrinsic properties of high-z galaxies to the IGM transmission of \lya\ photons and compared its predictions with several reionization histories proposed in recent studies \citep{Ishigaki2018, Finkelstein2019, Naidu2020, Yung2020b, Qin2025, Kageura2026}. The comparison with observed X$_{\mathrm{Ly}\alpha}$ evolution favored reionization histories in which the cosmic neutral fraction builds up progressively, while rapid transitions were difficult to reconcile with the data, primarily because of the relatively gradual decline of X$_{\mathrm{Ly\alpha}}$ observed between z = 6 and z = 8.

\end{itemize}

In this work, we demonstrated the key importance of the observed \lya\ emission line in constraining both IGM and galaxy evolution models from z = 4 to z = 14.2. Comparisons between the observed UV luminosity function and post-\jwst\ galaxy evolution models have shown that several different evolutionary scenarios can reproduce the available data \citep[e.g.,][]{Ferrara2023, Menci2024, Yung2024, Somerville2025}. The redshift evolution of \lya\ visibility provides an additional and independent diagnostic that may help break some of these model degeneracies, since the production and escape of \lya\ photons are linked to recent star formation episodes and ISM properties, and therefore change due to galaxy evolution at all redshifts.\\

Our finding that LAEs occupy a relatively stable region of galaxy-property space across the explored redshift range also motivates the identification of LAE candidates using photometric information alone \citep[e.g.,][]{Napolitano2023, Yoshioka2025, Vale2025}. Such approaches could enable the construction of substantially larger LAE candidate samples among SFGs over wide areas and from z = 0 to the high redshift frontier.
These samples could provide efficient target lists for wide-area \lya\ spectroscopic campaigns, including those performed with current facilities such as DESI \citep{Pinarski2026}, as well as future facilities including VLT/MOONS and ELT/MOSAIC.

Photometrically selected samples could also provide an empirical estimate of the evolution in the LAE fraction expected from galaxy evolution alone, in the absence of IGM attenuation. Comparing this intrinsic expectation with the observed X$_{\mathrm{Ly\alpha}}$ would offer an additional way to separate the effects of galaxy evolution from those of IGM transmission, thereby improving constraints on both the evolving galaxy population and the reionization history of the Universe.

Further progress will also require spectroscopic surveys capable of systematically measuring \lya\ velocity offsets in high-redshift galaxies. Ground-based facilities have provided such measurements up to z $\simeq$ 6.5 \citep[e.g.,][]{Erb2014, Cassata2020, Prieto-Lyon2025}. At higher redshift, these measurements require the higher spectral resolution provided by the \jwst/NIRSpec G140M and G140H dispersers, as demonstrated for a small number of sources \citep[e.g.,][]{Bunker2023B, Saxena2023B}. Expanding these measurements to statistically representative samples spanning a broad range of redshifts, UV absolute magnitudes, and galaxy properties will be essential for breaking the degeneracy between \lya\ velocity offsets and IGM transmission.
\begin{acknowledgements}
LN, LP, AA-P, AB, MB, S-JC, EC, VD, MG, KK, LP, CP, and RR acknowledge support from the ERC synergy grant 101166930 - RECAP. MC acknowledges support from the INAF GO Grant 2024 ”Revealing the nature of bright galaxies at cosmic dawn with deep JWST spectroscopy”. MLl acknowledges support from the INAF Mini-grant 2024 "Galaxies in the epoch of Reionization and their analogs at lower redshift", and the Large Grant RF 2023 F.O. 1.05.23.01.11 "The MOONS Extragalactic Survey". PS acknowledges financial support from INAF RF2024 Large Grant “UNDUST: UNveiling the Dawn of the Universe with JWST”.\\ 
This work is based on observations made with the NASA/ESA/CSA James Webb Space Telescope, obtained at the Space Telescope Science Institute, which is operated by the Association of Universities for Research in Astronomy, Incorporated, under NASA contract NAS5-03127. These observations are associated with programs \#1180, \#1181, \#1210, \#1211, \#1212, \#1213, \#1214, \#1215, \#1286, \#1287, \#1345, \#2198, \#2565, \#2750, \#3215, \#4106, \#4233, \#5224, \#6368, \#6585, and \#6541.
Support for program number GO-6368 was provided through a grant from the STScI under NASA contract NAS5-03127. The data were obtained from the Mikulski Archive for Space Telescopes (MAST) at the Space Telescope Science Institute. These observations can be accessed via \href{http://dx.doi.org/10.17909/0q3p-sp24}{DOI}.

Some of the data products presented in this work were retrieved from the Dawn JWST Archive (DJA). DJA is an initiative of the Cosmic Dawn Center (DAWN), which is funded by the Danish National Research Foundation under grant DNRF140.
\end{acknowledgements}

\bibliographystyle{aa}
\bibliography{biblio.bib}

\begin{appendix}

\section{A comparison between CAPERS and DJA reduced spectra} \label{sec:CAPERS_vs_DJA_app}
\begin{figure}[!ht]
\centering
\includegraphics[width=\linewidth]{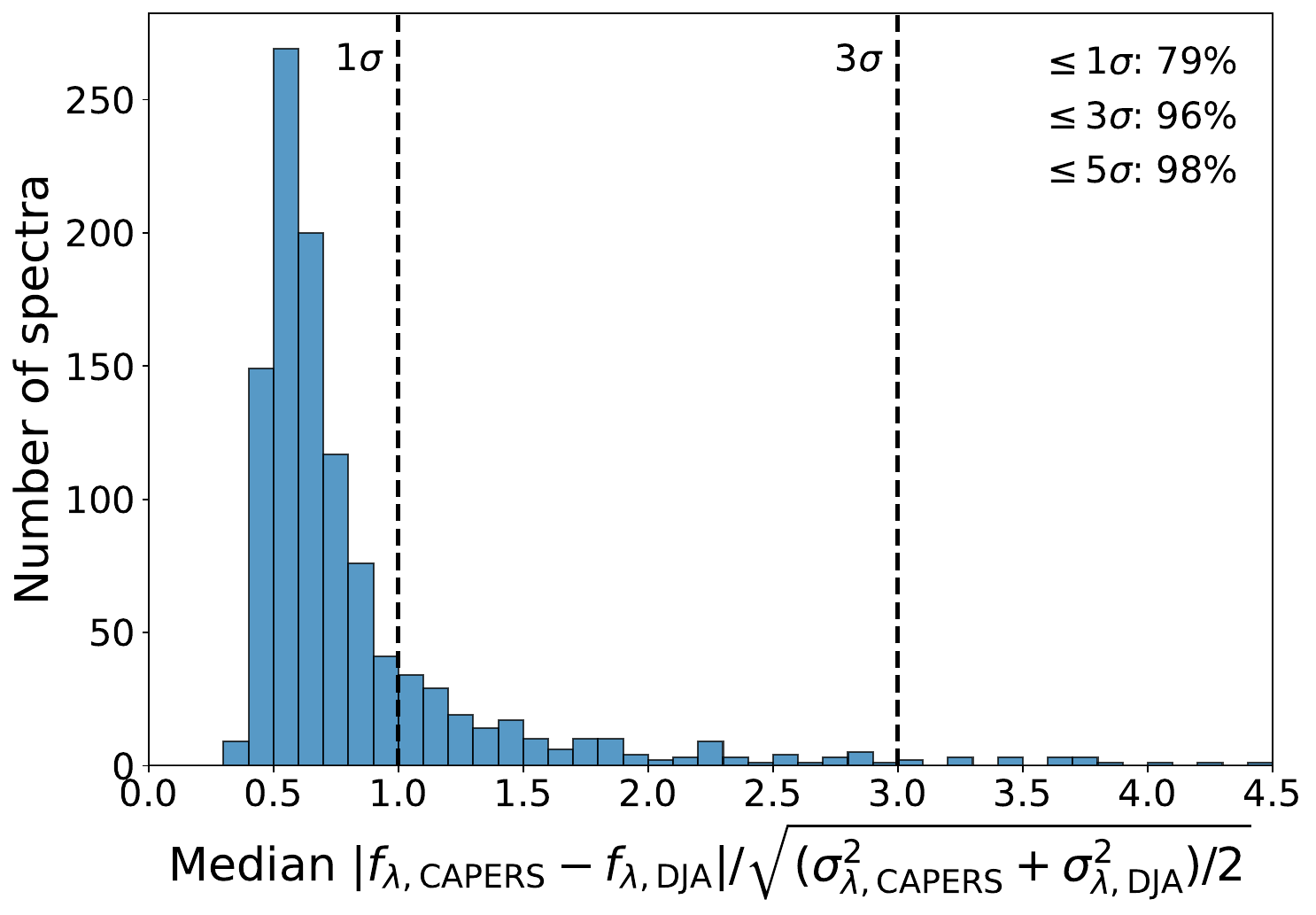}
\caption{Distribution of the median normalized absolute flux-density difference, $\delta_{\lambda}$, between the CAPERS and DJA reductions for the 1094 sources available in both data sets.}
\label{fig:CAPERSvsDJA_reduction}
\end{figure}

In this work, we combined spectra from two independent reduction frameworks: the internal CAPERS survey reduction, accounting for 36\% of the total sample, and the publicly available DJA reduction, accounting for the remaining 64\%. Both reductions are based on the official STScI JWST Calibration Pipeline, although they differ in some custom modifications adopted during the individual calibration and extraction steps (see Sect.~\ref{sec:Data_and_sample_selection} and references therein for details). To assess whether these differences could introduce systematic effects into our analysis, we directly compared the two reductions for targets available in both data sets, before applying any absolute spectro-photometric correction to the spectra (see Sect.~\ref{sec:CAPERS}). In total, 1094 galaxies at z > 4 have spectra independently reduced in both DJA version 4.4 and the internal CAPERS pipeline. For each source, we restricted the comparison to the wavelength range covered by both spectra and adopted the coarser of the two native wavelength samplings as the common grid, linearly interpolating the flux density ($f_{\lambda}$) and variance ($\sigma^2_{\lambda}$) of the other spectrum onto it. We then quantified the wavelength-dependent difference between the two reductions as:
\begin{equation}
    \delta_{\lambda} = \frac{|f_{\lambda \mathrm{, CAPERS}} - f_{\lambda \mathrm{, DJA}}|}{\sqrt{(\sigma^2_{\lambda \mathrm{, CAPERS}} + \sigma^2_{\lambda \mathrm{, DJA}})/2}}
\end{equation}
For the normalization, we adopted the root-mean-square uncertainty of the two reductions. This provides a more stringent consistency test than treating the two outputs as statistically independent, which would instead require adding their variances in quadrature. The latter choice would increase the denominator by a factor of $\sqrt{2}$ and therefore shift the entire $\delta_{\lambda}$ distribution toward values smaller by the same factor. We adopted the more conservative normalization because the CAPERS and DJA spectra originate from the same observations and are based on the official STScI JWST Calibration Pipeline: their reductions are therefore not expected to be fully statistically independent. 
For each galaxy, we computed the median $\delta_{\lambda}$ over all valid pixels within the common wavelength range. The resulting distribution for the population of 1094 galaxies is shown in Fig.~\ref{fig:CAPERSvsDJA_reduction}. The distribution is strongly concentrated below unity, with a median value of 0.6 and a 16--84th percentile range of 0.5--1.2. Overall, 79\%, 96\%, and 98\% of the spectra have median differences below $1\sigma$, $3\sigma$, and $5\sigma$, respectively. These results demonstrate that the CAPERS and DJA reductions are highly consistent relative to their quoted spectral uncertainties. To verify that no systematic differences affect the \lya\ EW$_0$ measurements specifically, we repeated the same calculation by restricting the comparison to the rest-frame wavelength range adopted for the \lya\ line fit (Sect.~\ref{sec:LyAmodel}). This test provides results fully consistent with those obtained over the unrestricted wavelength range. We therefore find no evidence that the use of the two different reduction frameworks introduces a significant systematic effect in our analysis, supporting their combination into the larger spectroscopic sample adopted throughout this work.

\section{Redshift dependence of the UV absolute magnitude completeness} \label{sec:Muv_completeness_app}
\begin{figure}[!ht]
\centering
\includegraphics[width=\linewidth]{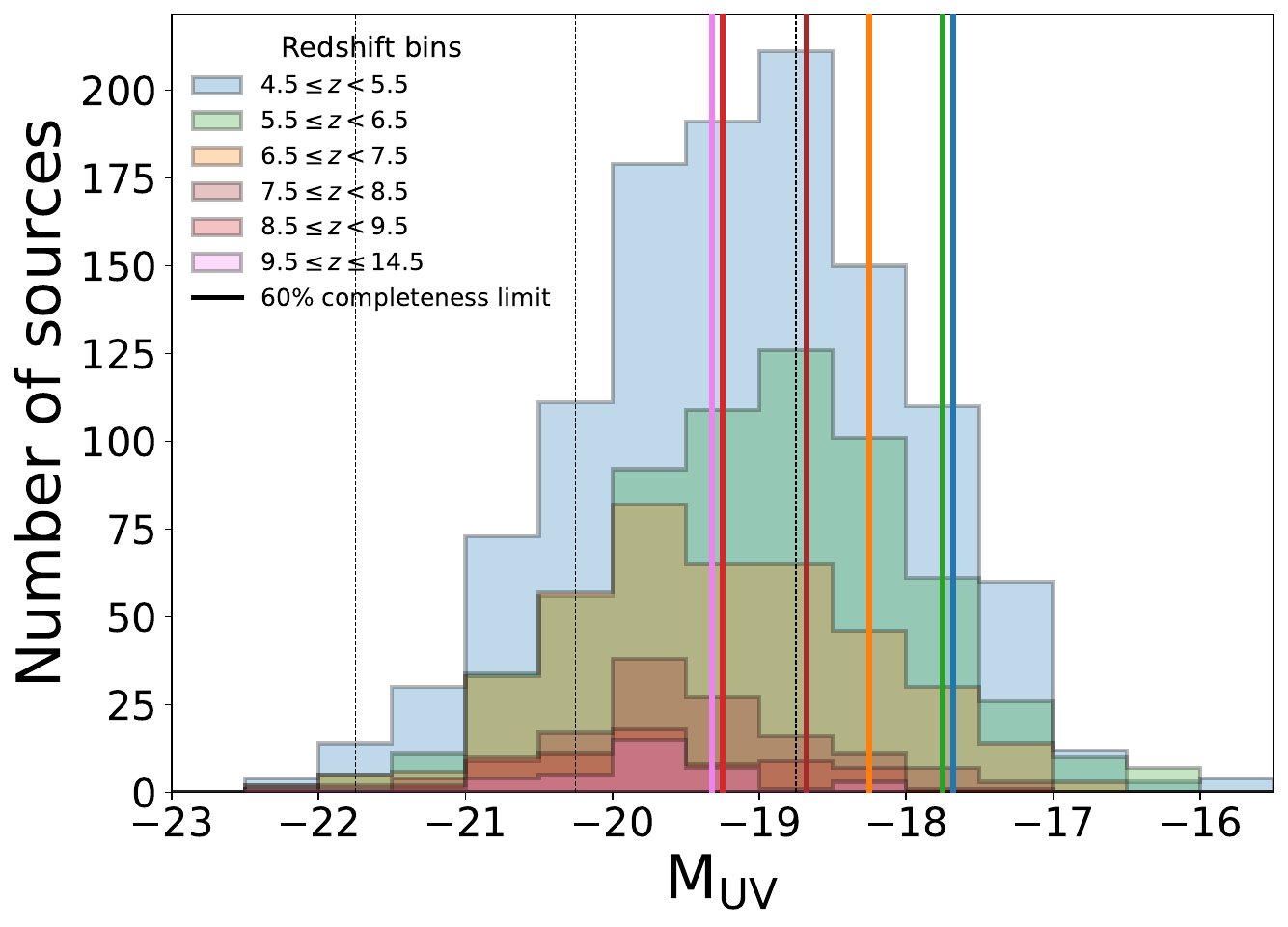}
\caption{Distribution of the observed M$_{\mathrm{UV}}$ of the considered sample in each redshift bin. Histograms and the corresponding 60\% UV absolute magnitude relative completeness limits are color coded by redshift. As in Fig.~\ref{fig:EW0compare}, the vertical dashed black lines mark M$_{\mathrm{UV}}$ = -21.75, –20.25, and –18.75, which delimit the UV-luminosity intervals adopted in our analysis.}
\label{fig:Muv_completenessVSz}
\end{figure}

Due to the flux-limited selection of observed targets in each considered JWST/NIRSpec spectroscopic surveys, we expect the observed samples to become progressively incomplete at the faint end toward higher redshifts. This observational bias can be quantified by measuring the redshift dependence of the UV absolute magnitude completeness limit, defined in Sect.~\ref{sec:Muv}. For each redshift bin adopted in our analysis (see Sect.~\ref{sec:lya_constraints}), namely z $\in$ [4.5, 5.5), [5.5, 6.5), [6.5, 7.5), [7.5, 8.5), [8.5, 9.5), and [9.5, 14.5), we derived the corresponding 60\% relative completeness limit in M$_{\mathrm{UV}}$. These limits are $\sim$ -17.75, -17.75, -18.25, -18.75, -19.25, and -19.25, respectively, and are shown in Fig.~\ref{fig:Muv_completenessVSz}. As expected, the completeness limit shifts toward brighter UV magnitudes at higher redshifts. \\
The bright UV magnitude interval of -21.75 < M$_{\mathrm{UV}}$ < -20.25 is well represented by the observed populations across the full redshift range considered. The intermediate interval adopted in our main analysis, -20.25 < M$_{\mathrm{UV}}$ < -18.75, remains above the 60\% relative completeness limit up to z $\leq$ 8.5, while its faint boundary becomes comparable to the completeness limits derived for the z $\geq$ 8.5 samples. Finally, we caution that population-level results in the faint UV-luminosity interval -18.75 < M$_{\mathrm{UV}}$ < -17.25, as defined by \cite{Kageura2025} for the investigation of the evolution of X$_{\mathrm{Ly\alpha}}$, are more strongly affected by incompleteness in unlensed JWST/NIRSpec field observations.

\section{The effect of assuming a constant $\xi_{\rm ion}$} \label{sec:AFM_const_xion}
\begin{figure*}[!ht]
\begin{minipage}{0.5\textwidth}
\centering
\includegraphics[width=\linewidth]{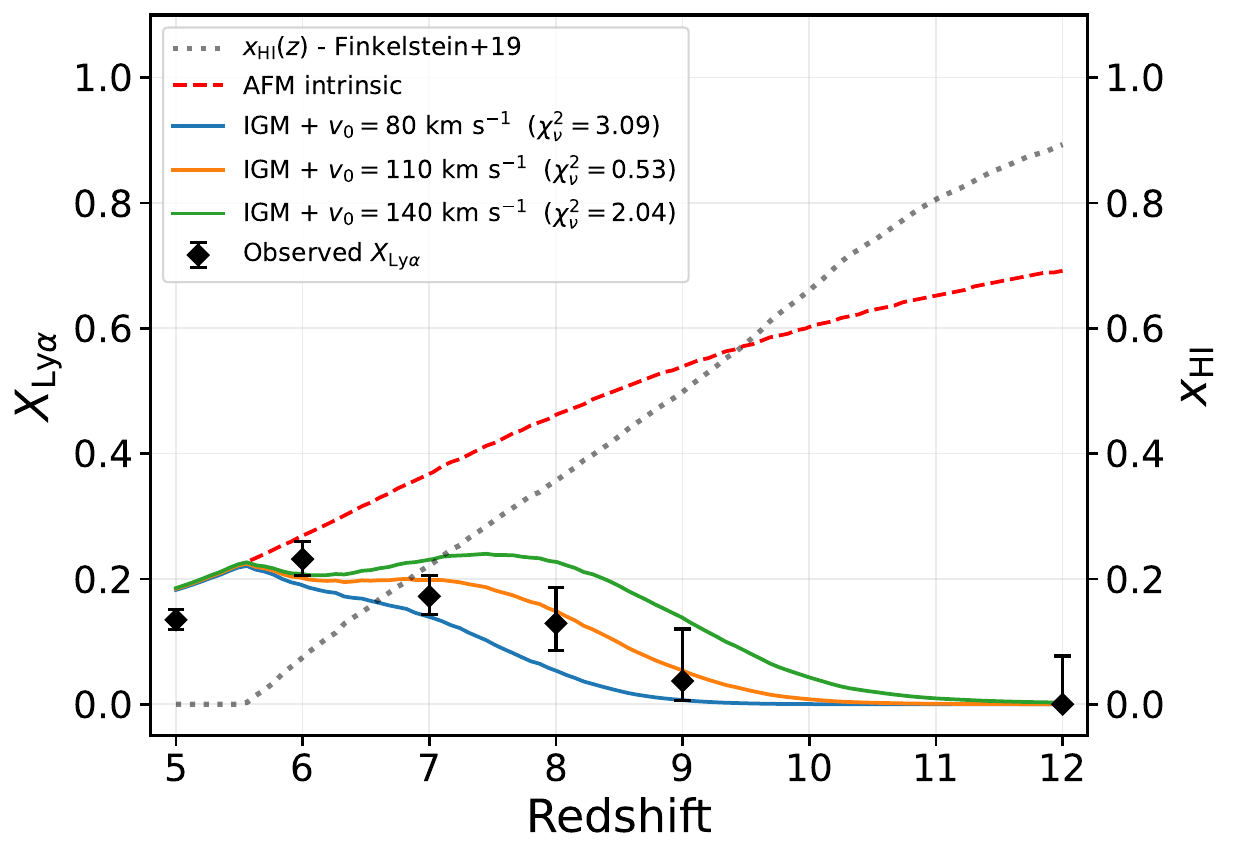}
\end{minipage}
\begin{minipage}{0.5\textwidth}
\centering
\includegraphics[width=\linewidth]{images/XLya_vs_redshift_Finkelstein19.pdf}
\end{minipage}
\caption{Left: Model prediction of X$_{\mathrm{Ly\alpha}}$ assuming the reionization history of \cite{Finkelstein2019} and a constant ionizing-photon production of $\log_{10} \xi_{\rm ion}$ = 25.2. Right: Model prediction of X$_{\mathrm{Ly\alpha}}$ assuming the reionization history of \cite{Finkelstein2019} and the redshift evolution of $\xi_{\rm ion}$ from \cite{Llerena2025}, as described in the main text. Symbols, colors, and line styles follow the same convention as in Fig.~\ref{fig:EoR_hystories}.}
\label{fig:Finkelstein19_xion_canon}
\end{figure*}
To investigate the impact of the assumed redshift evolution of the ionizing-photon production, we repeated the calculations described in Sect.~\ref{sec:AFM_pred} assuming instead a canonical constant value of $\log_{10} \xi_{\rm ion}$ = 25.2. We show the resulting predictions in Fig.~\ref{fig:Finkelstein19_xion_canon}, adopting the reionization history of \cite{Finkelstein2019} as reference. Compared with the fiducial redshift-dependent $\xi_{\rm ion}$ prescription adopted in Sect.~\ref{sec:reionization}, a constant $\xi_{\rm ion}$ produces a shallower intrinsic evolution of X$_{\mathrm{Ly\alpha}}$ driven by galaxy evolution alone. Consequently, the combined AFM and IGM predictions are shifted lower for a fixed v$_{\mathrm{0}}$, tending to favor slightly higher normalization values of the velocity offset.

\end{appendix}

\end{document}